\documentclass[apj]{emulateapj}
\pdfoutput=1
\usepackage{amsmath, natbib, graphicx, multirow}
\DeclareGraphicsExtensions{.pdf,.eps,.jpg,.png,.ps}

\newcommand{\asz}{\mbox{$A_\mathrm{SZ}$}}
\newcommand{\aprx}{\mbox{$\ensuremath{\sim}$}}
\newcommand{\ax}{\mbox{$A_\mathrm{X}$}}
\newcommand{\bsz}{\mbox{$B_\mathrm{SZ}$}}
\newcommand{\bx}{\mbox{$B_\mathrm{X}$}}
\newcommand{\csz}{\mbox{$C_\mathrm{SZ}$}}
\newcommand{\cx}{\mbox{$C_\mathrm{X}$}}
\newcommand{\da}{\mbox{$D_\mathrm{A}(z)$}}
\newcommand{\dainv}{\mbox{$D_\mathrm{A}^{-1}(z)$}}

\newcommand{\fx}{\mbox{$f_x$}}
\newcommand{\onefifty}{$150\,$GHz}
\newcommand{\LCDM}{\mbox{$\mathrm{\Lambda}$CDM}}
\newcommand{\Mfh}{\mbox{$M_{500}$}}

\newcommand{\MSZ}{\mbox{$M_{\mathrm{SZ}}^{500}$}}
\newcommand{\Mvir}{\mbox{$M_{\mathrm{vir}}$}}
\newcommand{\MX}{\mbox{$M_{\mathrm{X}}^{500}$}}
\newcommand{\msun}{\ensuremath{M_\odot}}

\newcommand{\rfh}{\mbox{$r_{500}$}}
\newcommand{\sqdeg}{\ensuremath{\mathrm{deg}^2}}
\newcommand{\tcmb}{\mbox{$T_\mathrm{CMB}$}}
\newcommand{\thcore}{\mbox{$\theta_\mathrm{c}$}}
\newcommand{\thcoresq}{\mbox{$\theta^2_\mathrm{c}$}}
\newcommand{\thint}{\mbox{$\theta_\mathrm{int}$}}

\newcommand{\uk}{\mbox{$\mu \mbox{K}$}}
\newcommand{\tzero}{\mbox{$\Delta T_0$}}
\newcommand{\Yfh}{\mbox{$Y_{\mathrm{SZ}}^{500}$}}

\newcommand{\Yrho}{\mbox{$Y_{\mathrm{SZ}}^{\rho}$}}
\newcommand{\Ysf}{\mbox{$Y_{\mathrm{SZ}}^{0.75'}$}}
\newcommand{\YSZ}{\mbox{$Y_{\mathrm{SZ}}$}}
\newcommand{\Yvir}{\mbox{$Y_{\mathrm{vir}}$}}
\newcommand{\Ytheta}{\mbox{$Y_{\mathrm{SZ}}^{\theta}$}}
\newcommand{\YthM}{\mbox{$Y_{\mathrm{SZ}}^{0.3\mathrm{Mpc}}$}}
\newcommand{\Ythresh}{\mbox{$Y_{\mathrm{SZ}}^{\mathrm{\phi}}$}(z)}
\newcommand{\YX}{\mbox{$Y_{\mathrm{X}}$}}

\begin{document}

\title{Measurement of Galaxy Cluster Integrated Comptonization \\ and Mass Scaling Relations with the South Pole Telescope}

\author{
B.~R.~Saliwanchik\altaffilmark{\CaseWestern}, 
T.~E.~Montroy\altaffilmark{\CaseWestern},
K.~A.~Aird\altaffilmark{\UChicago},
M.~Bayliss\altaffilmark{\Harvard,\CfA},
B.~A.~Benson\altaffilmark{\KICPChicago,\EFIChicago,\CPAFermilab},
L.~E.~Bleem\altaffilmark{\KICPChicago,\PhysicsUChicago,\ANL},
S.~Bocquet\altaffilmark{\Munich,\ExcellenceCluster},
M.~Brodwin\altaffilmark{\Miss},
J.~E.~Carlstrom\altaffilmark{\KICPChicago,\EFIChicago,\PhysicsUChicago,\ANL,\AAUChicago}, 
C.~L.~Chang\altaffilmark{\KICPChicago,\EFIChicago,\ANL}, 
H.~M. Cho\altaffilmark{\NIST}, 
A.~Clocchiatti\altaffilmark{\PUC},
T.~M.~Crawford\altaffilmark{\KICPChicago,\AAUChicago},
A.~T.~Crites\altaffilmark{\KICPChicago,\AAUChicago},
T.~de~Haan\altaffilmark{\McGill},
S.~Desai\altaffilmark{\Munich,\ExcellenceCluster},
M.~A.~Dobbs\altaffilmark{\McGill},
J.~P.~Dudley\altaffilmark{\McGill},
R.~J.~Foley\altaffilmark{\CfA,\illast,\illphy},
W.~R.~Forman\altaffilmark{\CfA},
E.~M.~George\altaffilmark{\Berkeley},
M.~D.~Gladders\altaffilmark{\KICPChicago,\AAUChicago},
A.~H.~Gonzalez\altaffilmark{\UFlorida},
N.~W.~Halverson\altaffilmark{\Colorado},
J.~Hlavacek-Larrondo\altaffilmark{\KIPAC,\Stanford},
G.~P.~Holder\altaffilmark{\McGill},
W.~L.~Holzapfel\altaffilmark{\Berkeley},
J.~D.~Hrubes\altaffilmark{\UChicago},
C.~Jones\altaffilmark{\CfA},
R.~Keisler\altaffilmark{\KICPChicago,\PhysicsUChicago},
L.~Knox\altaffilmark{\Davis},
A.~T.~Lee\altaffilmark{\Berkeley,\LBNL},
E.~M.~Leitch\altaffilmark{\KICPChicago,\AAUChicago},
J.~Liu\altaffilmark{\Munich,\ExcellenceCluster},
M.~Lueker\altaffilmark{\Berkeley,\Caltech},
D.~Luong-Van\altaffilmark{\UChicago},
A.~Mantz\altaffilmark{\KICPChicago},
D.~P.~Marrone\altaffilmark{\Arizona},
M.~McDonald\altaffilmark{\MIT},
J.~J.~McMahon\altaffilmark{\Michigan},
J.~Mehl\altaffilmark{\KICPChicago,\AAUChicago},
S.~S.~Meyer\altaffilmark{\KICPChicago,\EFIChicago,\PhysicsUChicago,\AAUChicago},
L.~Mocanu\altaffilmark{\KICPChicago,\AAUChicago},
J.~J.~Mohr\altaffilmark{\Munich,\ExcellenceCluster,\MPE},
S.~S.~Murray\altaffilmark{\CfA},
D.~Nurgaliev\altaffilmark{\Harvard}, 
S.~Padin\altaffilmark{\KICPChicago,\AAUChicago,\Caltech},
A.~Patej\altaffilmark{\Harvard},
C.~Pryke\altaffilmark{\Minnesota}, 
C.~L.~Reichardt\altaffilmark{\Berkeley},
A.~Rest\altaffilmark{\STScI},
J.~Ruel\altaffilmark{\Harvard},
J.~E.~Ruhl\altaffilmark{\CaseWestern}, 
A.~Saro\altaffilmark{\Munich},
J.~T.~Sayre\altaffilmark{\CaseWestern}, 
K.~K.~Schaffer\altaffilmark{\KICPChicago,\EFIChicago,\SAIC}, 
E.~Shirokoff\altaffilmark{\Berkeley,\Caltech}, 
H.~G.~Spieler\altaffilmark{\LBNL},
B.~Stalder\altaffilmark{\CfA},
S.~A.~Stanford\altaffilmark{\Davis,\LLNL},
Z.~Staniszewski\altaffilmark{\CaseWestern},
A.~A.~Stark\altaffilmark{\CfA}, 
K.~Story\altaffilmark{\KICPChicago,\PhysicsUChicago},
C.~W.~Stubbs\altaffilmark{\Harvard,\CfA}, 
K.~Vanderlinde\altaffilmark{\Dunlap,\Toronto},
J.~D.~Vieira\altaffilmark{\illast,\Caltech},
A.~Vikhlinin\altaffilmark{\CfA},
R.~Williamson\altaffilmark{\KICPChicago,\AAUChicago}, 
O.~Zahn\altaffilmark{\Berkeley,\LBNL,\BCCP},
A.~Zenteno\altaffilmark{\Munich,\ExcellenceCluster}
}

\altaffiltext{\CaseWestern}{Physics Department, Center for Education and Research in Cosmology and Astrophysics, Case Western Reserve University, Cleveland, OH 44106}
\altaffiltext{\UChicago}{University of Chicago,
5640 South Ellis Avenue, Chicago, IL 60637}
\altaffiltext{\Harvard}{Department of Physics, Harvard University, 17 Oxford Street, Cambridge, MA 02138}
\altaffiltext{\CfA}{Harvard-Smithsonian Center for Astrophysics,
60 Garden Street, Cambridge, MA 02138}
\altaffiltext{\KICPChicago}{Kavli Institute for Cosmological Physics, University of Chicago,
5640 South Ellis Avenue, Chicago, IL 60637}
\altaffiltext{\EFIChicago}{Enrico Fermi Institute, University of Chicago,
5640 South Ellis Avenue, Chicago, IL 60637}
\altaffiltext{\CPAFermilab}{Center for Particle Astrophysics, Fermi National Accelerator Laboratory, Batavia, IL 60510}
\altaffiltext{\PhysicsUChicago}{Department of Physics, University of Chicago,
5640 South Ellis Avenue, Chicago, IL 60637}
\altaffiltext{\ANL}{Argonne National Laboratory, 9700 S. Cass Avenue, Argonne, IL, USA 60439}
\altaffiltext{\Munich}{Department of Physics, Ludwig-Maximilians-Universit\"{a}t,
Scheinerstr.\ 1, 81679 M\"{u}nchen, Germany}
\altaffiltext{\ExcellenceCluster}{Excellence Cluster Universe,
Boltzmannstr.\ 2, 85748 Garching, Germany}
\altaffiltext{\Miss}{Department of Physics and Astronomy, University of Missouri, 5110 Rockhill Road, Kansas City, MO 64110}
\altaffiltext{\AAUChicago}{Department of Astronomy and Astrophysics, University of Chicago,
5640 South Ellis Avenue, Chicago, IL 60637}
\altaffiltext{\NIST}{NIST Quantum Devices Group, 325 Broadway Mailcode 817.03, Boulder, CO, USA 80305}
\altaffiltext{\PUC}{Departamento de Astronomia y Astrofisica, Pontificia Universidad Catolica, Chile}
\altaffiltext{\McGill}{Department of Physics, McGill University,
3600 Rue University, Montreal, Quebec H3A 2T8, Canada}
\altaffiltext{\illast}{Astronomy Department, University of Illinois at Urbana-Champaign,
1002 W.\ Green Street, Urbana, IL 61801 USA}
\altaffiltext{\illphy}{Department of Physics, University of Illinois at Urbana-Champaign,
1110 W.\ Green Street, Urbana, IL 61801 USA}
\altaffiltext{\Berkeley}{Department of Physics,
University of California, Berkeley, CA 94720}
\altaffiltext{\UFlorida}{Department of Astronomy, University of Florida, Gainesville, FL 32611}
\altaffiltext{\Colorado}{Department of Astrophysical and Planetary Sciences and Department of Physics, University of Colorado, Boulder, CO 80309}
\altaffiltext{\KIPAC}{Kavli Institute for Particle Astrophysics and Cosmology, Stanford University, 452 Lomita Mall, Stanford, CA 94305}
\altaffiltext{\Stanford}{Department of Physics, Stanford University, 452 Lomita Mall, Stanford, CA 94305}
\altaffiltext{\Davis}{Department of Physics, 
University of California, One Shields Avenue, Davis, CA 95616}
\altaffiltext{\LBNL}{Physics Division, Lawrence Berkeley National Laboratory,
Berkeley, CA 94720}
\altaffiltext{\Caltech}{California Institute of Technology, 1200 E. California Blvd., Pasadena, CA 91125}
\altaffiltext{\Arizona}{Steward Observatory, University of Arizona, 933 North Cherry Avenue, Tucson, AZ 85721}
\altaffiltext{\MIT}{Kavli Institute for Astrophysics and Space Research, Massachusetts Institute of Technology, 77 Massachusetts Avenue, Cambridge, MA 02139}
\altaffiltext{\Michigan}{Department of Physics, University of Michigan, 450 Church Street, Ann Arbor, MI, 48109}
\altaffiltext{\MPE}{Max-Planck-Institut f\"{u}r extraterrestrische Physik,
Giessenbachstr.\ 85748 Garching, Germany}
\altaffiltext{\Minnesota}{Physics Department, University of Minnesota, 116 Church Street S.E., Minneapolis, MN 55455}
\altaffiltext{\STScI}{Space Telescope Science Institute, 3700 San Martin
Dr., Baltimore, MD 21218}
\altaffiltext{\SAIC}{Liberal Arts Department, School of the Art Institute of Chicago, 
112 S Michigan Ave, Chicago, IL 60603}
\altaffiltext{\LLNL}{Institute of Geophysics and Planetary Physics, Lawrence
Livermore National Laboratory, Livermore, CA 94551}
\altaffiltext{\Dunlap}{Dunlap Institute for Astronomy \& Astrophysics, University of Toronto, 50 St George St, Toronto, ON, M5S 3H4, Canada}
\altaffiltext{\Toronto}{Department of Astronomy \& Astrophysics, University of Toronto, 50 St George St, Toronto, ON, M5S 3H4, Canada}
\altaffiltext{\BCCP}{Berkeley Center for Cosmological Physics, University of California, Berkeley, CA 94720}

\def\CaseWestern{1}
\def\UChicago{2}
\def\Harvard{3}
\def\CfA{4}
\def\KICPChicago{5}
\def\EFIChicago{6}
\def\CPAFermilab{7}
\def\PhysicsUChicago{8}
\def\ANL{9}
\def\Munich{10}
\def\ExcellenceCluster{11}
\def\Miss{12}
\def\AAUChicago{13}
\def\NIST{14}
\def\PUC{15}
\def\McGill{16}
\def\illast{17}
\def\illphy{18}
\def\Berkeley{19}
\def\UFlorida{20}
\def\Colorado{21}
\def\KIPAC{22}
\def\Stanford{23}
\def\Davis{24}
\def\LBNL{25}
\def\Caltech{26}
\def\Arizona{27}
\def\MIT{28}
\def\Michigan{29}
\def\MPE{30}
\def\Minnesota{31}
\def\STScI{32}
\def\SAIC{33}
\def\LLNL{34}
\def\Dunlap{35}
\def\Toronto{36}
\def\BCCP{37}

\email{benjamin.saliwanchik@case.edu}
\slugcomment{Submitted to \apj}

\begin{abstract}

We describe a method for measuring the integrated Comptonization (\YSZ) of clusters of galaxies from measurements of the Sunyaev-Zel'dovich (SZ) effect in multiple frequency bands and use this method to characterize a sample of galaxy clusters detected in South Pole Telescope (SPT) data.
We test this method on simulated cluster observations and verify that it can accurately recover cluster parameters with negligible bias.
In realistic simulations of an SPT-like survey, with realizations of cosmic microwave background anisotropy, point sources, and atmosphere and instrumental noise at typical SPT-SZ survey levels, we find that \YSZ\ is most accurately determined in an aperture comparable to the SPT beam size.
We demonstrate the utility of this method to measure \YSZ\ and to constrain mass scaling relations using X-ray mass estimates for a sample of 18 galaxy clusters from the SPT-SZ survey.
Measuring \YSZ\ within a $0.75'$ radius aperture, we find an intrinsic log-normal scatter of $21\pm11\%$ in \YSZ\ at a fixed mass.
Measuring \YSZ\ within a $0.3$ Mpc projected radius (equivalent to $0.75'$ at the survey median redshift $z = 0.6$), we find a scatter of $26\pm9\%$.
Prior to this study, the SPT observable found to have the lowest scatter with mass was cluster detection significance.
We demonstrate, from both simulations and SPT observed clusters, that \YSZ\ measured within an aperture comparable to the SPT beam size is equivalent, in terms of scatter with cluster mass, to SPT cluster detection significance.

\end{abstract}

\keywords{methods: data analysis --- galaxies: clusters --- X-rays: galaxies: clusters}

\section{Introduction}
\setcounter{footnote}{0}

Galaxy clusters are the largest gravitationally collapsed systems in the observed universe, 
and their abundance as a function of mass and redshift is a sensitive probe of the growth of structure in the universe.  
The ability to accurately and precisely estimate cluster masses is essential for using them to constrain cosmological parameters. 
Typically this is done through cluster observables, which do not directly measure cluster mass, but can be related to it through scaling relations \citep{vikhlinin09, rozo10, mantz10b, benson13}.
The Sunyaev-Zel'dovich (SZ) effect \citep{sunyaev72}, is caused by the inverse Compton scattering of cosmic microwave background (CMB) photons off of hot intra-cluster gas. It is a measure of the line-of-sight integral of the cluster pressure and is expected to be a low scatter proxy for cluster mass \citep{carlstrom02, kravtsov06a}.
In particular, the integrated Comptonization of a cluster, \YSZ, is expected to have a low intrinsic scatter with cluster mass and to be relatively insensitive to cluster astrophysics \citep{barbosa96, holder01a, motl05, nagai07, fabjan11}.

However, for SZ observations where the cluster size is on the order of the instrument beam size or smaller, there is typically a degeneracy in the constraints on the amplitude and shape of the assumed cluster profile \citep[e.g.,][]{benson04, planck11-5.1a, planck13-29}.
In this work, we present a Markov-Chain Monte Carlo (MCMC) analysis method for analyzing observations of the SZ effect, which measures \YSZ\ while marginalizing other SZ model parameters.
A feature of the MCMC method is that the \YSZ\ estimates it produces are well constrained even for clusters with relatively small radii on the sky.

We apply this MCMC method to simulated and real observations from the South Pole Telescope (SPT).
Previous analyses of clusters observed in the SPT-SZ survey used the cluster detection significance, $\xi$,  as a proxy for cluster mass \citep{vanderlinde10, andersson11, benson13, reichardt13}. 
Here we show that \YSZ\ integrated over a fixed angular aperture near the SPT beam size and $\xi$ have comparable fractional scatter in their respective mass scaling relations.  
\YSZ, however, is more easily compared to cluster parameters derived from other measurements.

\section{Cluster Sample and Observations}
\label{sec:clusters_and_obs}

\subsection{SZ Observations}
\label{sec:obs}

The South Pole Telescope is a 10-meter diameter off-axis Gregorian telescope with a 1 \sqdeg \ field of view, designed to operate at millimeter and submillimeter wavelengths \citep{carlstrom11}. 
In 2007-2011, the SPT surveyed 2500 \sqdeg \ in three frequency bands centered at 95, 150, and $220\,$ GHz. This survey is referred to as the SPT-SZ survey. The cluster sample used in this work is drawn from the two fields ($\aprx100$ \sqdeg \ each) observed with the SPT in 2008, one centered at right ascension (RA) $5^\mathrm{h}30^\mathrm{m}$, declination (Dec) $-55^\circ$ (J2000), and one at RA $23^\mathrm{h}30^\mathrm{m}$, Dec $-55^\circ$. A nearly identical cluster sample was used in \citet[][hereafter V10]{vanderlinde10}, \citet[][hereafter A11]{andersson11}, and \citet[][hereafter B13]{benson13}.

Observing procedures, data processing, and detection algorithms for these clusters are described in detail in V10 and \citet{staniszewski09}, and are summarized here. Details of the data processing pipeline are also described in \citet{schaffer11}.

Each field was observed by scanning the telescope back and forth in azimuth at $0.25^\circ/\mathrm{s}$, and then stepping in elevation and repeating until the entire field was covered. This process covers a 100 \sqdeg \ field in \aprx 2 hours. Field scans were repeated several hundred times until the noise in the co-added maps reached a completion depth of 18 \uk-arcmin for \onefifty. (See \citet{staniszewski09}, V10, or \citet{williamson11} for a description of field depth measurements.) The timestreams of the individual detectors were filtered to remove sky signal that was spatially correlated across the focal plane and long timescale detector drift. The combination of these filters effectively removes signals with angular scales larger than \aprx$0.5^\circ$. Data from individual detectors were combined using inverse-variance weighting, and the resulting maps were calibrated by comparison to the WMAP 5-year CMB temperature anisotropy power spectrum \citep{lueker10}.

\subsection{Cluster Detection}
Clusters are identified in the SPT maps using a matched filter (MF) \citep{haehnelt96, herranz02a, herranz02b, melin06}.
Specifics on this procedure can be found in \citet{staniszewski09} and V10 for single frequency cluster detection, and in \citet{williamson11} and \citet{reichardt13} for multi-frequency detection.
To locate clusters, the maps are multiplied in Fourier space with a filter matched to the expected spatial signal-to-noise profile of galaxy clusters. The matched filter, $\psi$, is given by: 
\begin{equation}
\psi(k_x,k_y) = \frac{B(k_x,k_y) S(|\vec{k}|)}{B(k_x,k_y)^2 N_{\mathrm{astro}}(|\vec{k}|) + N_{\mathrm{terr}}(k_x,k_y)},
\label{eq:psi} \end{equation}
where $B$ is the instrument response after timestream filtering, $S$ is the source template, and the noise has been divided into astrophysical ($N_{\mathrm{astro}}$), and terrestrial ($N_{\mathrm{terr}}$) components. $N_{\mathrm{astro}}$ includes power from primary and lensed CMB anisotropies, an SZ background from faint undetected clusters, and millimeter-wave emitting point sources. The noise power spectrum $N_{\mathrm{terr}}$ includes atmospheric and instrumental noise, estimated from jackknife maps. The source template is a two dimensional projection of an isothermal $\beta$-model, with $\beta$ set to 1 \citep{cavaliere76}:
\begin{equation} \label{eq:beta}
\Delta T = \Delta T_0 (1+\theta^2/\thcoresq)^{-1},
\end{equation}
where the central SZ temperature decrement $\Delta T_0$ and the core radius \thcore \ are free parameters.

Clusters are detected using a (negative) peak detection algorithm similar to SExtractor \citep{bertin96}. The significance of a detection, $\xi$, is defined to be the highest signal-to-noise (S/N) ratio across all \thcore.  In our analysis we use the unbiased significance,  $\zeta = \sqrt{\langle \xi \rangle^2 - 3}$, where $\langle \xi \rangle$ is the average detection significance of a cluster across many noise realizations (V10).

\subsection{Optical and X-ray Observations}
\label{sec:Xray_obs}

The optical and X-ray observations for the clusters used in this work have previously been described in A11 and B13, which we briefly describe here. 
All eighteen clusters have redshift measurements, fifteen of which are spectroscopic, and fourteen of the clusters have X-ray measurements. 

Optical $griz$ imaging and photometric redshifts for these clusters were obtained from the Blanco Cosmology Survey \citep{desai12}, and from pointed observations using the Magellan telescopes \citep{high10}. 
Of the fifteen clusters with spectroscopic redshifts, eight were obtained by the Low Dispersion Survey Spectrograph (LDSS3) on the Magellan Clay 6.5-m telescope \citep{high10}, and one by the Inamori Magellan Areal Camera and Spectrograph (IMACS) on the Magellan Baade 6.5-m telescope \citep{brodwin10}. 
The final six cluster redshifts were measured with IMACS and GMOS on Gemini South \citep{ruel13}. 
X-ray follow-up observations were performed with \emph{Chandra} ACIS-I and \emph{XMM-Newton} EPIC (A11, B13).

\section{MCMC Analysis Methods}
\label{sec:methods}

The application of MCMC methods to the detection and characterization of compact astrophysical sources in noisy backgrounds was proposed by \citet{hobson03}, and several experiments have used MCMC methods for parametrizing SZ signals from galaxy clusters. \citet{bonamente04}, \citet{bonamente06}, and \citet{laroque06} used MCMC methods to analyze SZ data from BIMA and OVRO, in conjunction with X-ray data from \emph{Chandra}, and fit $\beta$-model profiles to galaxy clusters. 
\citet{muchovej07}, \citet{culverhouse10}, and \citet{marrone09} parameterized SZA clusters, and \citet{halverson09} parameterized the Bullet Cluster using APEX-SZ data, all using the $\beta$-model. \citet{culverhouse10}, \citet{marrone09}, and \citet{marrone12} additionally estimated cluster \YSZ\ values.
Here we estimate galaxy cluster \YSZ\ values and \YSZ-$M$ scaling relations in addition to estimating $\beta$-model parameters.

\subsection{Posterior Distribution for a Compact Source}

We use a Metropolis-Hastings algorithm implementation of the Markov-Chain Monte Carlo method for parameter estimation.
For the case of a compact object with source template $S(\mathcal{H})$ in a two dimensional astronomical dataset $D$ with Gaussian noise, the likelihood has the form:
\begin{equation}
\mathrm{P}(D|\mathcal{H}) = \frac{\mathrm{exp}(-\frac{1}{2}[D-S(\mathcal{H})] C^{-1} [D-S(\mathcal{H})]^*)}{(2\pi)^{N_{\mathrm{pix}}/2} |C|^{1/2}},
\label{eq:Pr_1band} \end{equation}
where $C$ is the noise covariance matrix for the dataset $D$, and $N_{\mathrm{pix}}$ is the number of pixels in $D$ \citep{hobson03}. In this method, $C$ is composed of the combined $N_{\mathrm{astro}}$ and $N_{\mathrm{terr}}$ noise terms in the matched filter $\psi$ (equation \ref{eq:psi}). 

We are interested in parametrizing galaxy clusters using the SZ effect, which is the spectral distortion they produce in the blackbody CMB spectrum. At two of the SPT's observing frequencies (95 and \onefifty) this distortion is manifested as a decrement in CMB power, while the net change in CMB power at 220 GHz is negligible.

Equation \ref{eq:Pr_1band} is easily generalizable to the case of astronomical images in multiple frequency bands, where the 
unnormalized log likelihood may be calculated in the Fourier domain as \\ \\
$ \begin{array}{@{\hspace{0mm}}r@{\;}l@{\hspace{0mm}}}
\mathrm{Log} \left( \overline{\mathrm{P}}(D|\mathcal{H}) \right) =
\end{array} $
\begin{equation} -\frac{1}{2}\sum_{\bar{k},\nu_i,\nu_j}\frac{\left(\widetilde{D}_{\nu_i}(\bar{k})-\widetilde{s}^{\mathcal{H}}_{\nu_i}(\bar{k})\right)\left(\widetilde{D}_{\nu_j}(\bar{k})-\widetilde{s}^{\mathcal{H}}_{\nu_j}(\bar{k})\right)^*}{N_{\nu_i \nu_j}(\bar{k})},
\end{equation}
where $\widetilde{D}_{\nu_i}(\bar{k})$ is the Fourier transform of the map for frequency $\nu_i$, $\widetilde{s}^{\mathcal{H}}_{\nu_i}$ is the frequency dependent Fourier transform of the cluster model for parameter set $\mathcal{H}$, and $N_{\nu_i\nu_j}(\bar{k})$ is the frequency dependent covariance matrix for the $\nu_i$ and $\nu_j$ frequency maps. Here $N_{\nu_i\nu_j}(\bar{k})$ is simply the multiband extension of the covariance matrix $C$ in equation \ref{eq:Pr_1band}.

\subsection{Implementation}

Our MCMC is modeled after the generic Metropolis-Hastings method described in \citet{hobson03}, and is implemented in MATLAB \footnote{Mathworks Inc., Natick MA, 01760}. 

In this work, we use the MCMC method for cluster parametrization, not detection. Our testing found that it was more computationally costly and not more effective at cluster detection than the MF method.  
Throughout this work, our MCMC is run over a relatively small area of sky (512 pixels $\times$ 512 pixels, or \aprx $2^\circ \times 2^\circ$) centered on a cluster which has already been identified. 

Cluster parameter recovery is tested in single and multi-frequency simulations below (\S \ref{sec:params}), but we use only \onefifty \ when investigating scaling relations (for observed and simulated clusters) to match the SPT cluster analysis in B13, from which our sample is derived. We use the $\beta$-model source template given in equation \ref{eq:beta}.
\citet[][in preparation]{montroy13} demonstrate, using simulations and methods similar to those described in \S \ref{sec:params}, that \YSZ\ is recovered accurately with a $\beta$-model for either $\beta$-model or Arnaud profile \citep{arnaud10} input clusters.  

Clusters are characterized by four parameters: their location on the sky in RA and Dec, the magnitude of the SZ temperature decrement \tzero, and the core radius \thcore. We apply priors in the form of uniform probability distributions in each parameter. Given that we are characterizing clusters that have already been detected by the MF, our position priors can be quite tight. We impose a simple square-box prior on RA and Dec, centered at the MF cluster location and extending $\pm 1.25'$. Our \tzero \ and \thcore \ priors restrict these parameters to broadly reasonable values given the expected mass and redshift range of our cluster sample. Our SZ temperature decrement prior is $-2.5~\mathrm{mK} \le \tzero \le 0.0~\mathrm{mK}$, and our radius prior is $0.025' \le \thcore \le 2.5'$. \thcore \ is not allowed to fall to zero for numerical reasons.

\begin{figure*}
\clearpage
\plotone{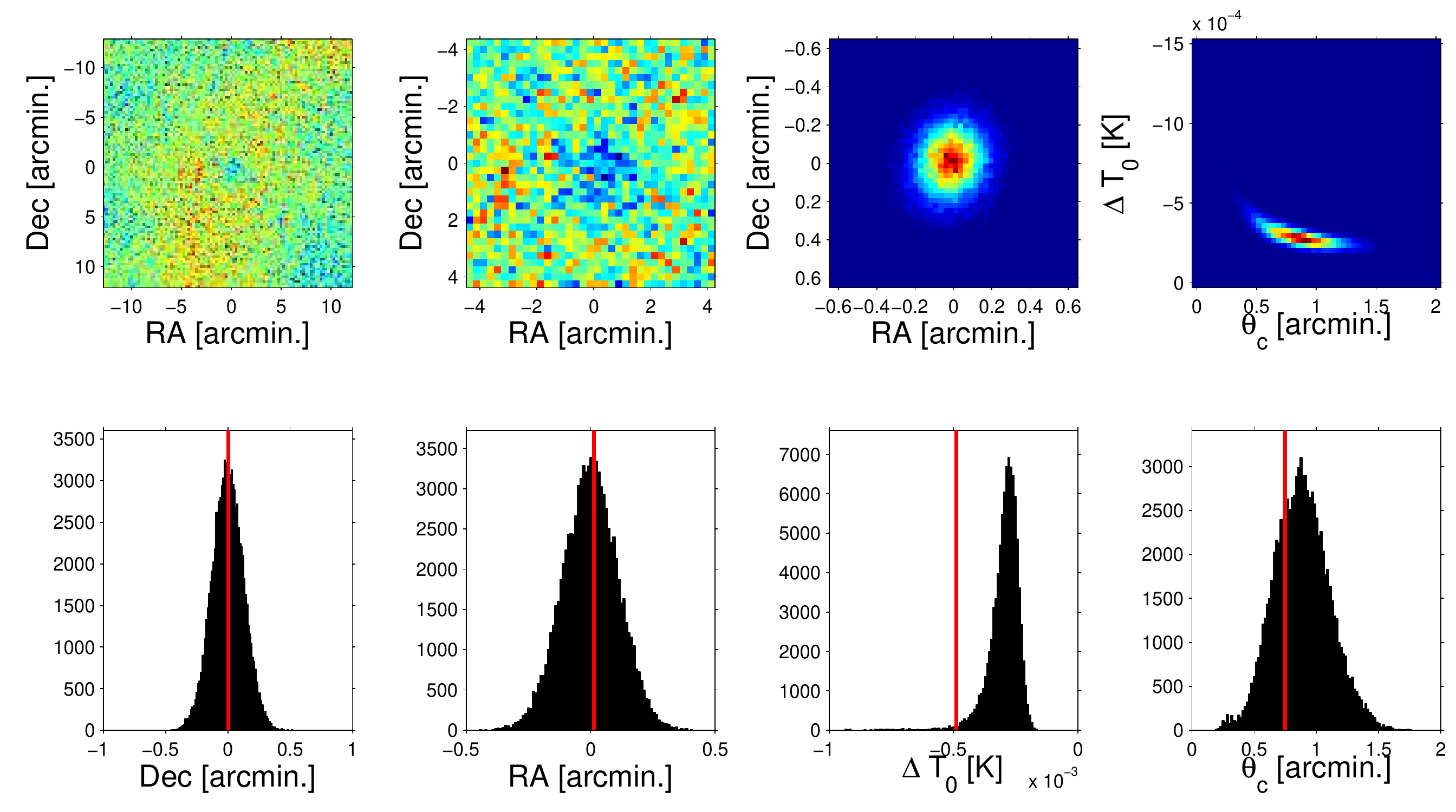}
\caption{From left to right the top row shows (1) a $25' \times 25'$ section of an SPT sky map centered on cluster SPT-CL J$2341$-$5119$ ($\xi = 9.65$, $z = 0.9983$) (2) a close up of $7.5' \times 7.5'$ centered on the cluster location, (3) a 2D histogram of the cluster position values from the MCMC chain, marginalizing over \tzero \ and \thcore, and (4) a 2D histogram of \tzero \ and \thcore, marginalizing over position. Likewise the bottom row shows one dimensional marginalized histograms of the parameters (5) Declination (6) Right Ascension, (7) \tzero, and (8) \thcore. 
Vertical red lines in the bottom row indicate the matched filter parameter values for this cluster.
}
\label{fig:param_dist}
\end{figure*}

Burn-in, as evaluated by stability of the likelihood values, is typically complete within several hundred steps. For the 12,000 simulated cluster realizations in \S \ref{sec:params} we cut the first $10^3$ steps, using the rest of the $10^4$ steps to characterize the probability surface. 
In the scaling relation analysis discussed in \S \ref{sec:scaling} many fewer clusters were analyzed, allowing the chain length to be extended to $10^5$ steps, from which we exclude the first  $10^4$ steps in order to ensure convergence. We define recovered parameter values to be the median of the MCMC equilibrium distribution for each parameter, marginalizing over the other parameters.  Uncertainties are given by the $68\%$ confidence interval of the marginalized distribution for each parameter, centered on the median value. Figure \ref{fig:param_dist} shows the parameter distributions for a typical cluster detected with the SPT.

\section{Simulations}
\label{sec:sims}

\subsection{Simulated Thermal SZ Cluster Maps}

We used two sets of simulations; one uses $\beta$-model clusters (defined by \tzero \ and $\thcore$) to investigate cluster parameter recovery (\S \ref{sec:params}), while the second uses cluster gas profiles inferred from dark matter light cone simulations to calibrate \YSZ-$M$ scaling relations (\S \ref{sec:scaling}). The second set of simulations is described in detail in \citet{shaw10}, and will be referred to as the S10 simulations for convenience. The thermal SZ (tSZ) cluster profiles used in each set are discussed in more detail in the relevant sections below.

\subsection{Astrophysical Backgrounds}
\label{sec:astro}

We use simulated maps of astrophysical backgrounds that include contributions from the CMB and extragalactic point sources. 
Simulated CMB anisotropies were generated based on realizations of the gravitationally lensed WMAP 5-year \LCDM \ CMB power spectrum.

The extragalactic point source population at \onefifty \ consists of two classes of objects: ``dusty'' sources dominated by thermal dust emission from star formation bursts, and ``radio'' sources dominated by synchrotron emission.  We use the source count model of \citet{negrello07} at 350 GHz, which is based on physical modeling by \citet{granato04} for dusty sources. Source counts at \onefifty \ are estimated by assuming the flux densities scale as $S_{\nu} \propto \nu^\alpha$, where $\alpha = 3$ for high-redshift protospheroidal galaxies, and $\alpha = 2$ for late-type galaxies. For radio sources we use the \citet{dezotti05} model at \onefifty, which is in agreement with observed radio source populations \citep{vieira10, mocanu13}. 

Point source population realizations were generated by sampling from Poisson distributions for each population in bins with fluxes from 0.01 mJy to 1000 mJy. Sources were randomly distributed across the map. Correlations between sources or with galaxy clusters were not modeled, following V10.
These \onefifty \ simulated point source populations were used for the scaling relation simulations of \S \ref{sec:scaling}, but not for the multiband pipeline checks of \S \ref{sec:params}.

\subsection{Simulated Observations}
\label{sec:instr}

We model the SPT transfer function for the 95 GHz and 150 GHz frequency bands by producing synthetic timestreams from simulated maps convolved with the SPT beam, observing them using the SPT scan strategy, and convolving the resulting timestreams with detector time constants. We produce maps by performing data processing, as in \S \ref{sec:obs}, on the simulated timestreams. To simplify the complex computational task of processing large sky maps, the transfer function was modeled as a 2D Fourier filter. V10 shows that this approximation introduces systematic errors in the recovered cluster $\xi$ values of less than 1\%.

The instrumental and atmospheric noise in the SPT maps were estimated by creating difference maps, which were constructed to have no astrophysical signal.
Each field consists of several hundred individual observations.
We randomly multiply half of the observations by -1, and then coadd the full set of observations.
We repeat this several hundred times, each time calculating the two-dimensional spatial power spectrum, which we average to estimate the instrumental and atmospheric noise in the coadded SPT map.
This averaged noise spectrum is used to generate random map realizations of the SPT noise, which are added to the simulated maps.

\section{Pipeline Checks}
\label{sec:params}

\subsection{Cluster Model}
\label{sec:TM_sims}

We use mock observations of clusters in simulated sky maps to evaluate the accuracy and bias of the recovered cluster parameters.
We begin with simulated maps that contain the astrophysical signals described in Section \ref{sec:astro}.
To this we add mock clusters with an assumed $\beta$-model profile, with known SZ decrements and radii, at specified locations.
Simulated SPT observations are then performed on these maps.  
Three different cluster core radii ($0.25'$, $0.5'$, and $1.0'$) are used, combined with eight values for peak Comptonization between 175 \uk \ and 2 mK, spanning the range of values typically found for SPT-detected clusters with $\xi > 5$. 
These cluster profiles are convolved with the SPT transfer function, and then placed in the simulated maps.
For each combination of $\beta$-model cluster parameters we create five pairs of simulated maps (\onefifty \ and 95 GHz) by placing 100 copies of the cluster at random locations in five unique noise maps. 
This results in 500 noise realizations for each combination of cluster parameters, or 12,000 clusters total. 
As usual, \aprx $2^\circ \times 2^\circ$ cutouts are made around each cluster, and the MCMC is run on each patch. 

In \S \ref{sec:fund_parms} and \S \ref{sec:YSZ} we test parameter recovery in the single-band (150 GHz) and multiband (95 GHz and 150 GHz) cases.
In this paper, we do not use the 220 GHz SPT measurements, because they are not deep enough to make significant improvements to the CMB subtraction.

\subsection{Position, Radius, and Amplitude}
\label{sec:fund_parms}

\begin{figure}
\includegraphics[width=\columnwidth]{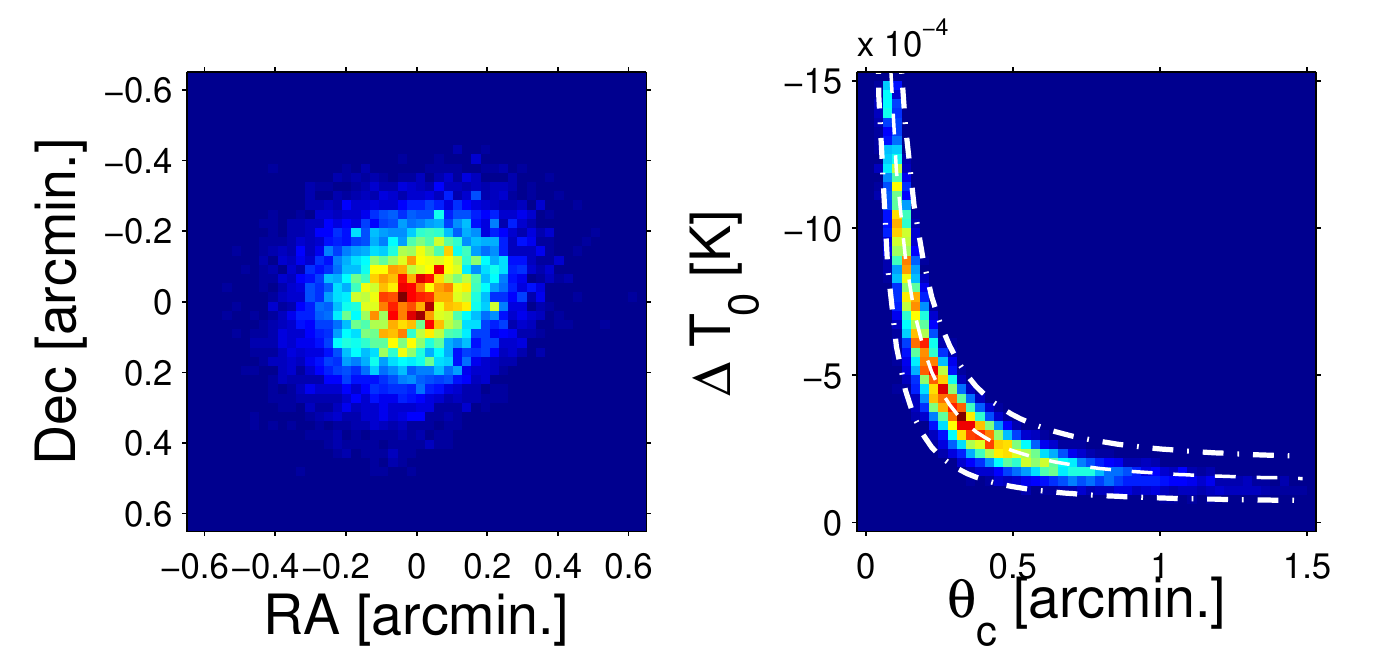}
\caption{SPT-CL J$0533$-$5005$ ($\xi = 5.59$, $z = 0.8810$), an SPT observed cluster with core radius $\thcore < 1'$. The left figure shows a 2D histogram of the cluster position, and the right figure shows \tzero\ and \thcore. For clusters near the SPT beam size ($\aprx1'$ FWHM at \onefifty) and selection threshold, the position is well constrained, however the radius and amplitude are degenerate.  Despite this, the integrated Comptonization, \YSZ, is well constrained. The over-plotted curves are \YSZ\ iso-curves. The dashed line is the recovered \YSZ\ for this cluster, while the dot-dashed lines are $\pm50\%$ \YSZ.}
\label{fig:degen}
\end{figure}

We first examine the recovered values of the four baseline cluster parameters: the right ascension (RA) and declination (Dec) position, \tzero, and \thcore.  
The cluster positions are measured accurately, and we find no bias in either position parameter.
For clusters near the SPT beam size ($\aprx1'$ FWHM at \onefifty) and selection threshold, the amplitude and shape of the cluster will not be well constrained, however the integrated signal within the SPT beam will be.
A similar degeneracy has previously been noted in other cluster analyses \citep[e.g.,][]{benson04, planck11-5.1a, planck13-29}.
In Figure \ref{fig:degen}, we show the recovered cluster parameters for a typical cluster in the SPT catalog (SPT-CL J$0533$-$5005$, $\xi = 5.59$, $z = 0.8810$, $\thcore < 1.0'$).
While the position is well-constrained, there is a significant degeneracy between the constraints on \thcore\ and \tzero.

\subsection{Integrated Comptonization}
\label{sec:YSZ}

In general, the integrated Comptonization of a cluster is calculated by integrating the source function, $S(\theta)$, out to a given angular aperture \thint:
\begin{equation}
\YSZ = 2 \pi \int_0^{\thint}S(\theta) \ \theta \ \mathrm{d}\theta.
\label{eq:ysz_theta}
\end{equation}
For much of this work, \thint \ will be a constant angular aperture. We distinguish this estimator of \YSZ\ from others by referring to it as \Ytheta \ hereafter.

In the case of a two dimensional projection of a spherical $\beta$-model with $\beta=1$ (equation \ref{eq:beta}), this integral can be solved analytically:
\begin{equation}
\Ytheta = \frac{\pi \tzero \thcoresq}{\fx \tcmb} \ \mathrm{Log} \left[1+\left(\frac{\thint}{\thcore}\right)^2\right],
\label{eq:y_theta}
\end{equation}
where \thcore\ is the core radius in arcminutes, \tzero\ is the central temperature decrement in units of $K_\mathrm{CMB}$, the equivalent CMB temperature fluctuation required to produce the observed power fluctuation, \tcmb\ is the CMB blackbody temperture of 2.725~K, and \fx\ is given by:
\begin{equation}
\fx = \left( x \frac{e^x+1}{e^x-1} - 4 \right) \left[ 1+\delta(x,T_e) \right],
\end{equation} 
where $x = h \nu / k \tcmb$, and $\delta(x,T_e)$ accounts for relativistic corrections to the SZ spectrum \citep{itoh98,nozawa00}. For the details of the calculation of \fx\ for the SPT see A11.

We use this equation to calculate \Ytheta \ for every step in the MCMC chain, and thus to produce a marginalized distribution of \Ytheta \ values.
In these simulations, integration to a radius approximately corresponding to the \onefifty\ SPT beam diameter (roughly the range $0.75' < \thint < 1.25'$) produces \Ytheta \ distributions that are well constrained despite the degeneracy of \thcore \ and \tzero, with minimal error in recovered cluster \Ytheta \ values.  
Integration in this section is performed to \thint \ $= 0.75'$, though other values are explored for scaling relations in \S \ref{sec:scaling} below. Note that we calculate \Ytheta\ from the marginalized distribution of the source model parameters, not by integrating the flux on the sky.

If the redshift of a cluster is known it is also possible to integrate \YSZ\ within an angular aperture corresponding to a specific physical radius, $\rho$:
\begin{equation}
\thint = \rho \ \dainv, \nonumber
\end{equation} 
where \da\ is the angular diameter distance to the redshift $z$. In Sections \ref{sec:scaling} and  \ref{sec:spt_clusters}, we examine \YSZ\ integrated within a constant physical radius, $\rho$, for all clusters in a sample. We will refer to this quantity as as \Yrho.

\begin{figure}
\includegraphics[width=\columnwidth]{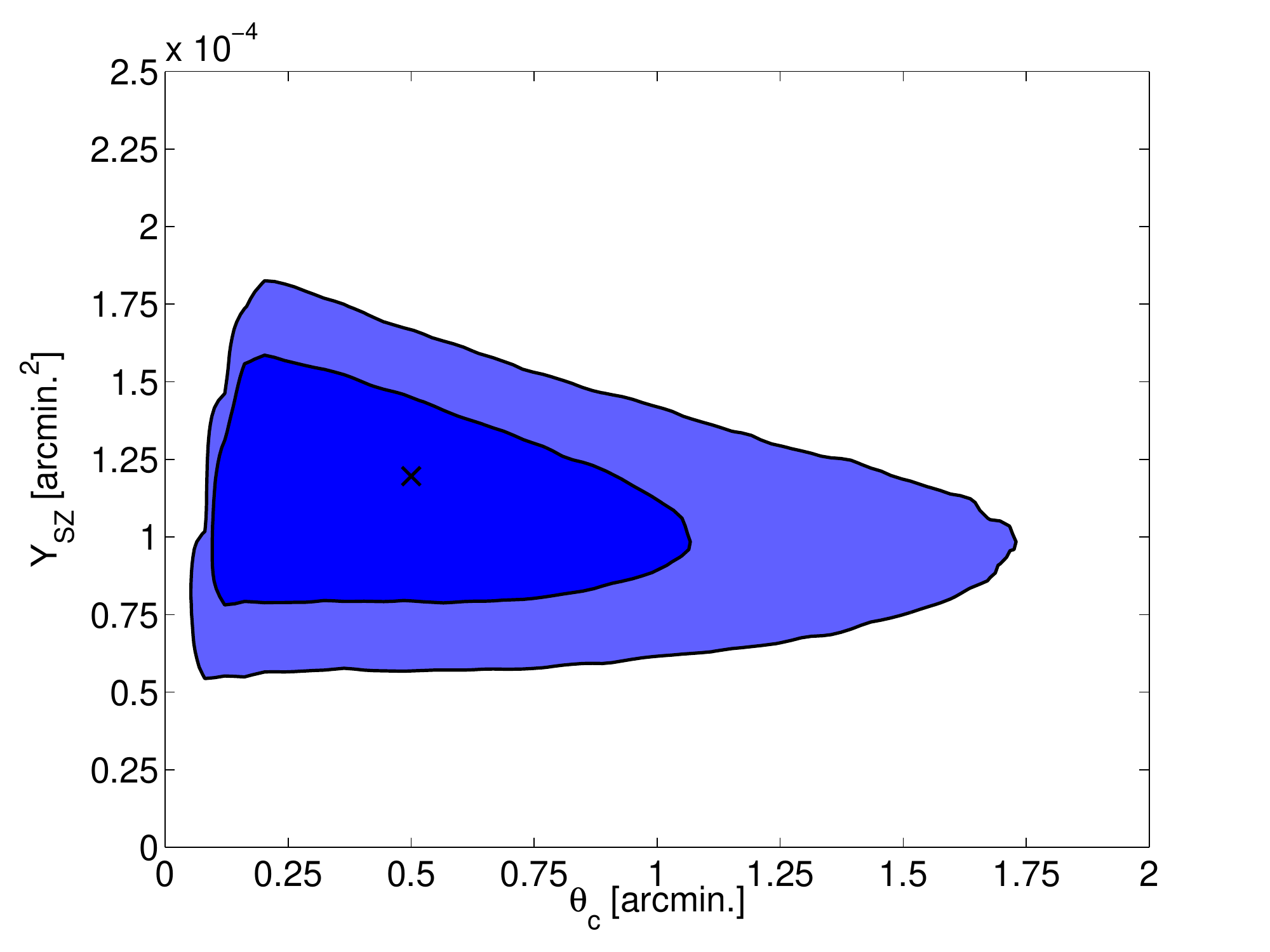}
\caption{The marginalized constraints on \YSZ\ and \thcore\ from 500 noise realizations of a typical simulated cluster with radius smaller than the SPT \onefifty \ beam ($\thcore = 0.5'$, $\tzero = 300 \mu K$, $\xi = 6.2$). The contours show the $68\%$ and $95\%$ confidence regions. The `X' marks the input \thcore\ and \YSZ\ values ($\Ysf = 1.20 \times 10^{-4}$ arcmin.$^2$). Despite only having an upper bound on \thcore, \YSZ\ is well constrained.}
\label{fig:ysz_vs_theta}
\end{figure}

In Figure \ref{fig:degen} we show a typical SPT cluster in which \Ytheta\ is well constrained despite the degeneracy between \tzero\ and \thcore.
Figure \ref{fig:ysz_vs_theta} shows \YSZ\ and \thcore\ parameter distributions for 500 runs of a typical simulated cluster with a radius smaller than the SPT \onefifty\ beam size ($\thcore = 0.5'$, $\tzero = 300 \mu K$, $\xi = 6.2$). 
The cutoff at low \thcore\ is due to the small, but non-zero, minimum priors on \thcore\ and \tzero, this is not a feature of the data likelihood.  
Despite only having an upper bound on \thcore, \YSZ\ is still well constrained.

\begin{figure}
\includegraphics[width=\columnwidth]{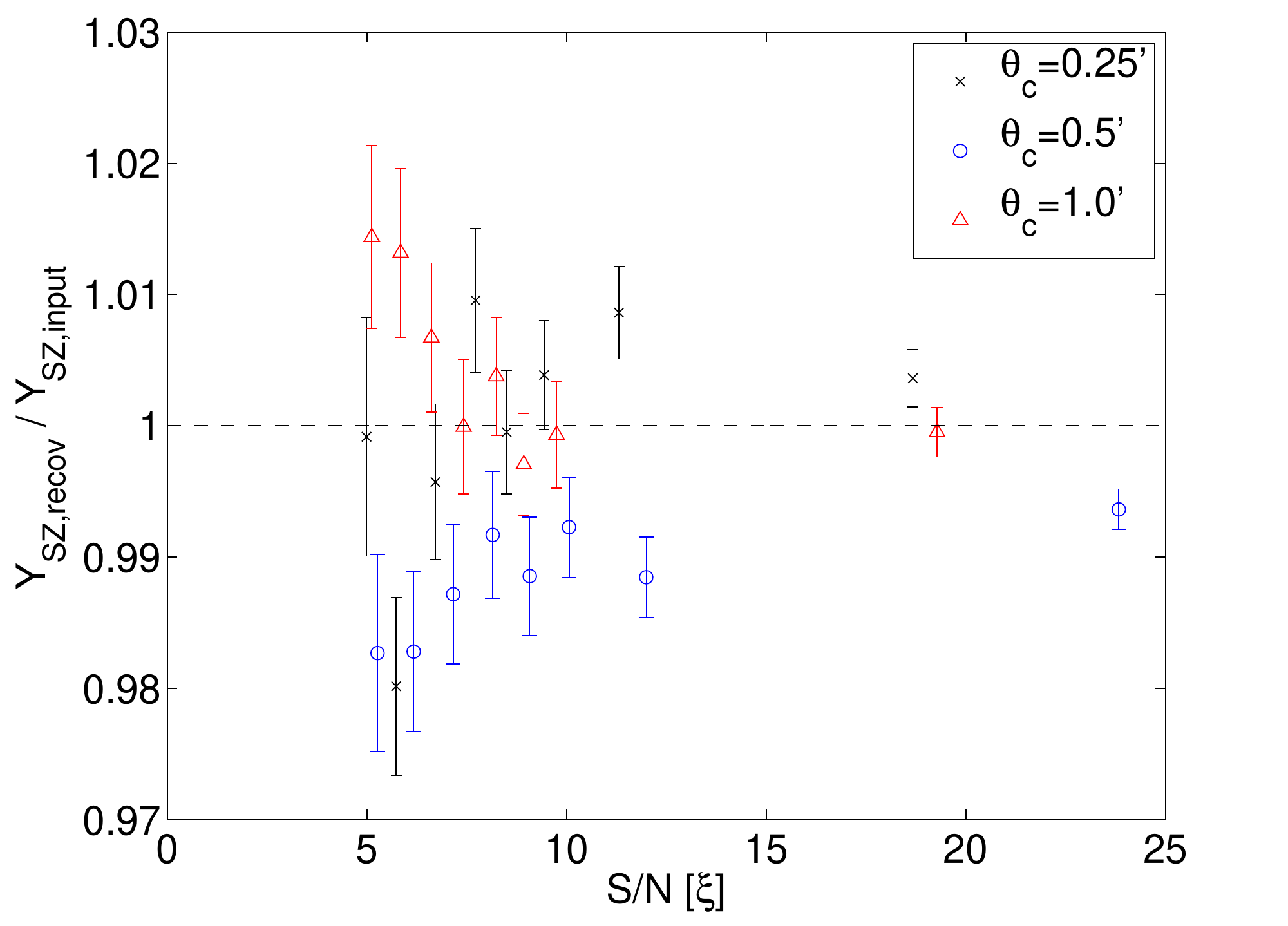}
\caption[]{Average ratios of recovered to input \YSZ\ for 24 different combinations of \thcore \ and \tzero. Each point is the mean recovered \YSZ\ for a simulated cluster with 500 independent noise realizations. The errorbars represent the error on the mean of these recovered \YSZ\ values.}
\label{fig:Y_ratio}
\end{figure}

In Figure \ref{fig:Y_ratio}, we show the ratio of the recovered to input \YSZ\ as a function of core radius and cluster detection significance, $\xi$, for 24 different combinations of \thcore \ and \tzero, each with 500 independent noise realizations.
Despite a slight apparent bias for some \thcore\ values, we find no significant bias as a function of the detection significance, and recover Ysz accurately to $<2\%$ in all cases.
On average recovered \Ytheta \ values are $0.27\%$ lower than input values, which is below the $0.49\%$ error in the mean.

\section{Scaling Relations from Simulated Clusters}
\label{sec:scaling}

In this section, we compare \YSZ\ and $\zeta$ as SZ observables for the SPT-SZ survey, focusing on their scatter with cluster mass.
To do this, we use maps derived from the S10 simulations, which are intended to provide more realistic cluster profiles than the $\beta$-model clusters used in \S \ref{sec:params}.

The steepness of the galaxy cluster mass function will introduce bias in a scaling relation fitted in the presence of noise or intrinsic scatter in the population. Therefore, in \S \ref{sec:tsz_sims} we fit \Ytheta-$M$ scaling relations for clusters in simulated tSZ-only maps, to minimize the selection bias. These maps contain none of the celestial or instrumental noise spectra described in \S \ref{sec:sims} \ (CMB, point sources, atmospheric noise, and instrumental noise), only tSZ signal. In \S \ref{sec:sky_sims} we fit for a \Ytheta-$M$ scaling relation using clusters in S10 simulation maps containing the full astrophysical and instrumental noise terms to evaluate the performance of the MCMC in the presence of noise.

\subsection{Simulated Clusters}
\label{sec:shaw_sims}
 
The S10 simulations are based on a dark matter lightcone simulation, with cosmological parameters consistent with the WMAP 5-year data and large-scale structure measurements \citep{dunkley09}. To include baryons in the simulations, \citet{shaw10} apply the semi-analytic gas model of \citet{bode07}, specifically their fiducial model, to the dark matter halos identified in the output of the lightcone simulation. From the simulations, we construct two dimensional SZ intensity maps at \onefifty \ of clusters with virial mass (\Mvir) greater than $5 \times 10^{13} \msun h^{-1}$ by summing the electron pressure density along the line of sight. The resulting maps are projections of all the clusters in the lightcone simulation onto a simulated sky. Forty $10^\circ \times 10^\circ$ maps were produced by this procedure, together with catalogs of cluster masses, redshifts, and positions.  

\subsection{\YSZ-$M$ Scaling Relation Fitting Methods}
\label{sec:var_theta}

We assume a scaling between \YSZ\ and $M$ of the form:
\begin{equation}
\YSZ = \asz \left(\frac{\Mvir}{3 \times 10^{14} \, \msun h^{-1}}\right)^{\bsz} \left(\frac{E(z)}{E(0.6)}\right)^{\csz},
\label{eq:scaling_rel} \end{equation}
parametrized by the normalization \asz, the mass scaling \bsz, and the redshift evolution \csz, and where $E(z) \equiv H(z)/H_0$.
For self-similar evolution, $\bsz = 5/3$ and $\csz = 2/3$ (e.g., \citet{kravtsov06a}). 
The pivot points of the scaling relation were defined to match the approximate mean mass and redshift for the SPT cluster sample.  

We fit the \Ytheta-\Mvir\ scaling relation by minimizing the fractional scatter, $\mathcal{S}$, in \Ytheta, defined as:
\begin{equation}
\mathcal{S} = \sqrt{\frac{1}{N} \displaystyle\sum_{n=1}^N \left( \frac{Y^{\mathrm{recov}}_{n}-Y^{\mathrm{input}}_{n}}{Y^{\mathrm{input}}_{n}} \right)^2},
\label{eq:frac_scatter} \end{equation}
where $Y^{\mathrm{recov}}_{n}$ is the integrated Comptonization recovered by the MCMC for the $n^{th}$ cluster, $Y^{\mathrm{input}}_{n}$ is the corresponding Comptonization calculated from the input catalog mass and the assumed scaling relation (equation \ref{eq:scaling_rel}), and we sum over $N$ simulated clusters. The scaling relation parameters \asz, \bsz, and \csz \ are varied using a grid search method, and the scatter $\mathcal{S}$ is calculated at each point in the parameter space. The combination of parameters that minimizes $\mathcal{S}$ is taken to be the best-fit set of parameters. This definition of fractional scatter is used to fit \Ytheta-\Mvir\ scaling relations in \S \ref{sec:tsz_sims} and in \S \ref{sec:sky_sims}.

\subsection{Results for Simulated Thermal-SZ-Only Maps}
\label{sec:tsz_sims}

We run both the MCMC and MF methods on tSZ-only maps from the S10 dark matter lightcone simulations described in \S \ref{sec:shaw_sims}. 
These simulated tSZ maps contain only thermal SZ signal, and no CMB, point sources, atmospheric noise, or instrumental noise.

We measure the SZ signal in these maps using the methods described in Sections \ref{sec:TM_sims} and \ref{sec:YSZ} for clusters with $\Mvir > 4 \times 10^{14} \msun h^{-1}$, and redshift $0.3 < z < 1.2$. 
We then use the cluster virial masses and equation \ref{eq:scaling_rel} to find the best fit scaling relation parameters by minimizing the fractional scatter in equation \ref{eq:frac_scatter}. We do this for both the \YSZ-$\Mvir$ and $\zeta$-$\Mvir$ scaling relations, which allows for direct comparison of these analysis methods. 
The redshift range corresponds to the redshift range of observed SPT clusters, and the mass criteria corresponds to the mass of clusters at the lower SPT significance limit of the \citet{reichardt13} cluster catalog ($\xi = 4.5$), at the survey median redshift of $z = 0.6$. 

As a baseline for the scatter in the measured \YSZ-\Mvir\ scaling relations for these simulations, we examine the intrinsic scatter between \Mvir\ and \Yvir, the contribution to the SZ flux from within the spherical virial radius for each cluster.
We fit the \Yvir-\Mvir\ scaling relation parameters using the same method as for measured \YSZ\ values, and find the fractional scatter in the best-fit scaling relation to be $16\%$. 

\begin{figure*}
\clearpage
\plotone{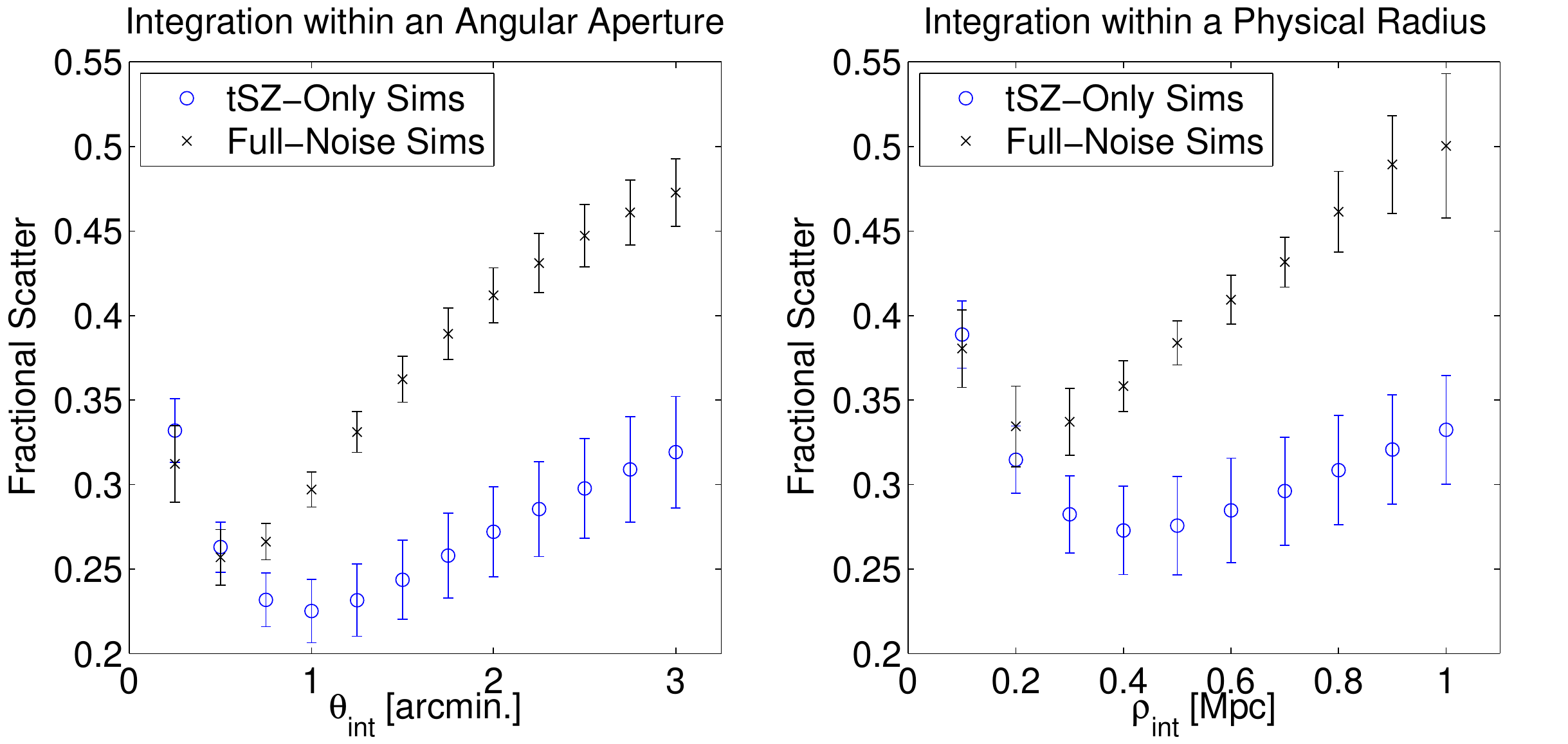}
\caption[]{Fractional scatter vs. integration radius for tSZ-only and full noise simulations. Left panel: fractional scatter vs. integration angle in arcminutes. Right panel: fractional scatter vs. integration radius in megaparsecs. The scatter in the tSZ-only simulations is essentially the intrinsic scatter in the population, since only tSZ fluctuations are present. Adding the other noise terms shifts the scatter up, and the minimum down in angular or physical scale because those noise terms dominate at large angles. The optimal angular apertures correspond roughly to the optimal physical radii at the median redshift of the cluster sample, $z = 0.6$.}
\label{fig:frac_scatter}
\end{figure*}

We fit \Ytheta-$\Mvir$ relations for a range of angular apertures, $\theta$, with \Ytheta\  defined in equation \ref{eq:y_theta}. 
Figure \ref{fig:frac_scatter} shows the fractional scatter as a function of the integration angle for angles ranging from $0.25'$ to $3.0'$. 
We find that the fractional scatter in \Ytheta \ does not vary significantly with angular aperture, with a broad minimum in the scatter at $\aprx 0.75'$ - $1.0'$ (\Ysf). 
The exact location of the minimum scatter shifts between the tSZ-only maps and the full-noise S10 maps, but is near $0.75'$ in both cases (see Figure \ref{fig:frac_scatter}).
For simplicity, and for comparison between the different simulated maps and observed clusters, we use the \Ysf-\Mvir\ scaling relation as our nominal scaling relation.
The \Ysf-$\Mvir$ scaling relation has $23\pm2\%$ fractional scatter in \YSZ, which is slightly less than the $27\pm2\%$ scatter in the $\zeta$-$\Mvir$ scaling relation for these clusters.  
The scatter in the tSZ-only simulations is primarily due to the intrinsic scatter in the mass to SZ observable scaling, scatter from the tSZ background is sub-dominant.
(Note, the scatter here is fractional scatter, whereas previous SPT analyses in V10 and B13 quoted a log-normal scatter, at a level consistent with the values found in this work.)
 
\begin{figure}
\includegraphics[width=\columnwidth]{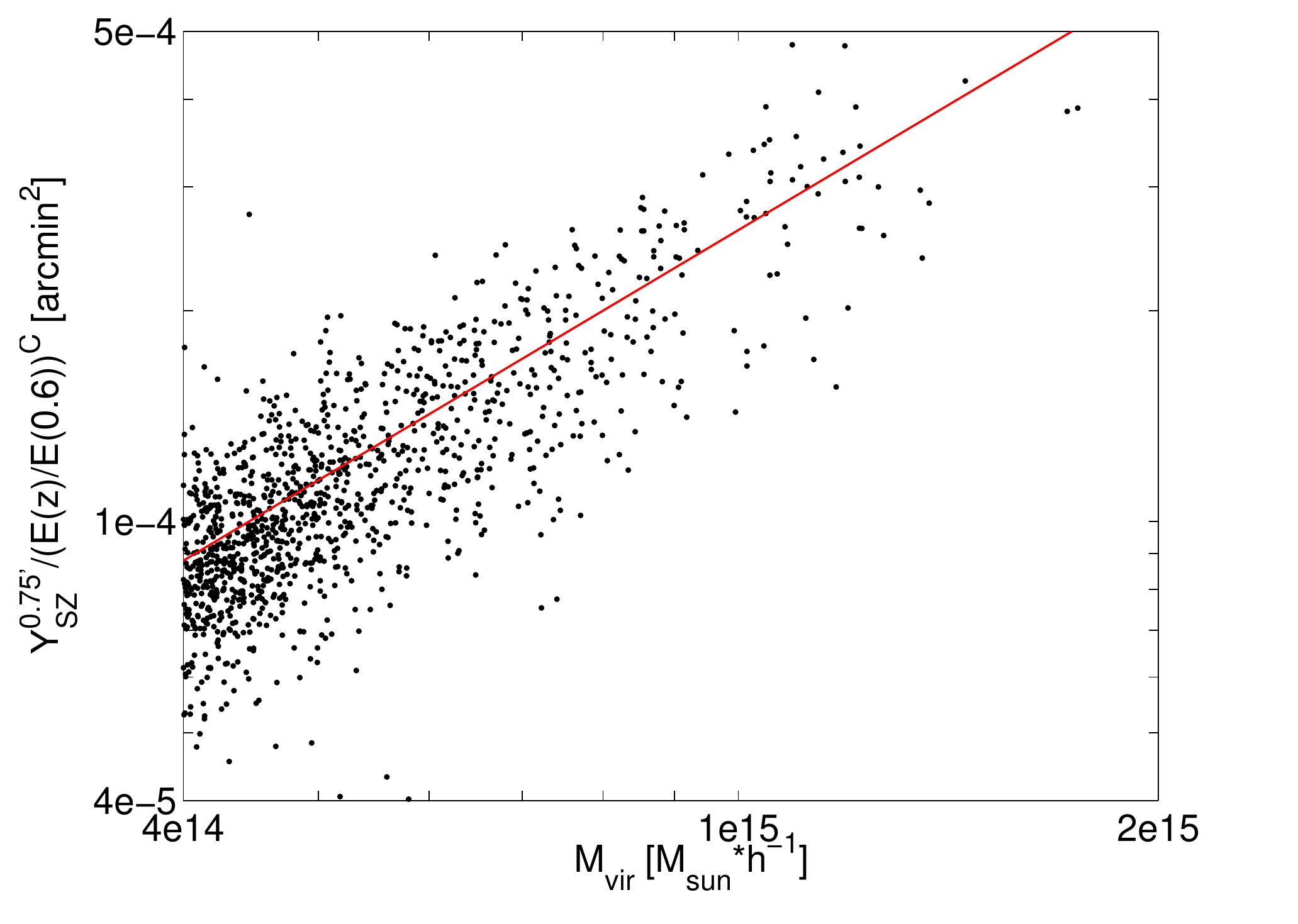}
\caption[]{\Ysf \ versus $\Mvir$ \ for 1187 mass-selected clusters in the S10 simulated tSZ-only maps, where we only include clusters with $\Mvir > 4 \times 10^{14} \msun h^{-1}$ in the redshift range $0.3 < z < 1.2$. Fractional scatter in \YSZ\ is $23\pm2\%$. The solid line is the best-fit \Ysf-$\Mvir$ scaling relation found for this cluster sample.}
\label{fig:shaw_sz}
\end{figure}

Figure \ref{fig:shaw_sz} shows \Ysf \ versus $\Mvir$ for the 1187 clusters examined from this simulation. The solid line is the best-fit \Ysf-$\Mvir$ scaling relation found for these clusters.  The scaling relation parameters (\asz, \bsz, \csz, and $\mathcal{S}$) for the \Ysf \ scaling relation are given in Table \ref{tab:scaling_rels}.

We also calculate \YSZ\ within a constant physical radius, $\rho$, (\Yrho) for all the clusters in the catalog. 
The angular size of a cluster is a function of its redshift, therefore, it is interesting to measure \YSZ\ within a fixed physical radius.
In Figure \ref{fig:frac_scatter}, we plot the best-fit scatter for a range of integration radii between 0.1 to 1.0 Mpc. 
We find that the minimum fractional scatter in \YSZ\ within a fixed physical radius is higher than the minimum fractional scatter within a fixed angular aperture.
For \Yrho, the scatter is increased by the varying angular size of the chosen physical radius at different redshifts. 
The optimal physical radius corresponds roughly to the optimal angular aperture, at the median redshift of the cluster sample, $z = 0.6$. 
Clusters farther from the median redshift will have integration angles farther from the optimal angle, resulting in relatively higher scatter in \Yrho\ than in \Ytheta.

As can be seen in Figure \ref{fig:frac_scatter}, we find a broad minimum in scatter at $\aprx 0.3$ - $0.4$ Mpc, with a minimum scatter of $27\pm3\%$. This is comparable to the $\zeta$-$\Mvir$ relation for these clusters, and slightly higher than the scatter in the \Ysf-$\Mvir$ scaling relation. The scaling relation parameters for \YSZ\ within $0.3$ Mpc (\YthM), ($0.3$ Mpc being equivalent to $0.75'$ at $z = 0.6$) are given in Table~\ref{tab:scaling_rels}.

\subsection{Results for Full-Noise Simulated Maps}
\label{sec:sky_sims}

We also fit \Ytheta-$\Mvir$ scaling relations for the simulated clusters in the presence of other astrophysical and instrumental noise components (see \S \ref{sec:astro} and \S \ref{sec:instr}). The same cluster sample ($\Mvir > 4 \times 10^{14} \msun h^{-1}$, and $0.3 < z < 1.2$) was analyzed in this set of simulations as in the simulated tSZ-only maps. We will refer to this set of simulations as the full-noise S10 simulated maps.

The scaling relation fitting for the clusters from this set of simulations was performed as in \S \ref{sec:tsz_sims}. As in \S \ref{sec:tsz_sims}, the scatter is a weak function of angular aperture, with the minimum shifted to $\aprx 0.5'$ - $0.75'$. Figure \ref{fig:frac_scatter} shows the fractional scatter as a function of angular aperture of integration. 

For the \Ysf-$\Mvir$ scaling relation we find a fractional scatter in \YSZ\ of $27\pm1\%$. Since the scatter here includes both intrinsic scatter and the measurement uncertainty, we expect it to be larger than the scatter in \Ysf \ in \S \ref{sec:tsz_sims}. This level of scatter is comparable to the $27\pm2\%$ scatter in $\zeta$ found in the $\zeta$-$\Mvir$ scaling relation for these simulations. 

\begin{figure}
\includegraphics[width=\columnwidth]{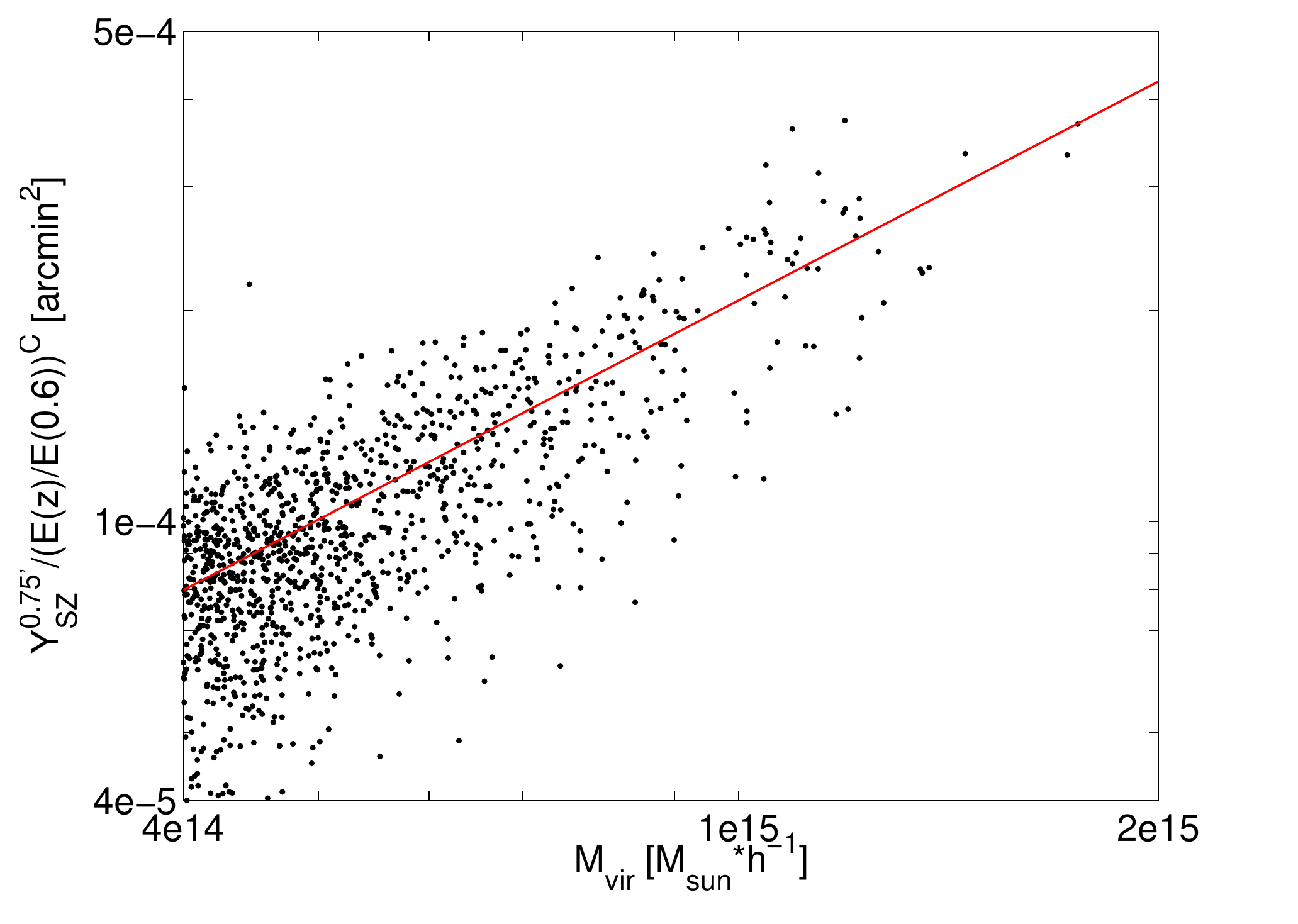}
\caption[]{\Ysf \ versus $\Mvir$ for 1187 mass-selected clusters in the full-noise S10 simulations, which include CMB, point sources, astrophysical noise, and realistic SPT instrument noise. We include only clusters with $\Mvir > 4 \times 10^{14} \msun h^{-1}$ in the redshift range $0.3 < z < 1.2$. Fractional scatter in \YSZ\ is $27\pm1\%$. The solid line is the best-fit \Ysf-$\Mvir$ scaling relation found for this cluster sample.}
\label{fig:shaw_sky}
\end{figure}

Figure \ref{fig:shaw_sky} shows \Ysf \ versus $\Mvir$ for the 1187 clusters analyzed from the full-noise S10 simulated maps. The solid line is the best-fit \Ysf-$\Mvir$ scaling relation found for this cluster sample. The mass scaling relation parameters for \Ysf\ in this set of simulations are given in Table \ref{tab:scaling_rels}.

Using these simulations we also calculate \Yrho\ for a range of $\rho$ values, as in \S \ref{sec:tsz_sims}, and fit \Yrho-$\Mvir$ scaling relations for each $\rho$. 
Figure \ref{fig:frac_scatter} shows the fractional scatter as a function of the integration radius for a range of physical radii. 
We find a broad minimum in scatter at $\aprx0.2$ - $0.3$ Mpc, with a minimum scatter of $33\pm2\%$.
The optimal integration radius shifts down here relative to the simulated tSZ-only maps because of the scale dependence of the noise sources added in the full-noise S10 maps, which dominate the scatter in these simulations. 
In particular, the noise induced by the simulated CMB increases with angular scale, leading to a preference for smaller integration radii. 
The scatter in \Yrho\ for these simulations is slightly higher than the scatter in both the $\zeta$ and the \Ysf\ mass scaling relations. 
The scaling relation parameters for the nominal \YthM\ mass scaling relation are given in Table \ref{tab:scaling_rels}. 
The optimal physical radius again corresponds roughly to the optimal angular aperture, at the median redshift of the cluster sample.

\section{\YSZ\ for SPT Observed Clusters}
\label{sec:spt_clusters}

\subsection{\YSZ-\Mfh\ Scaling Relation Fitting Methods}
\label{sec:spt_clusters_scaling}

In this section we perform \Ytheta-$M$ scaling relation fitting for a sample of SPT observed clusters, using the same scaling relation as in the simulations (equation \ref{eq:scaling_rel}) and the X-ray determined cluster masses. 
In this section we define cluster mass as \Mfh, the mass inside a spherical radius \rfh, within which the average density is 500 times the critical density of the universe at the cluster's redshift.
To fit this scaling relation with clusters selected in the SPT-SZ survey, we have to account for the shape of the cluster mass function and the SPT survey selection, which was based on the SPT significance, $\xi$. 
This is similar to the procedure followed in previous SPT analyses \citep[B13, V10]{reichardt13}, with the added complication that in this work we must express the SPT selection function in \YSZ\ instead of $\xi$.

The (unnormalized) probability of a mass $M$ given an integrated Comptonization \YSZ\ is given by:
\begin{equation}
\overline{\mathrm{P}}(M|\YSZ) = \mathrm{P}(\YSZ|M)\mathrm{P}(M), \nonumber
\end{equation}
where $\mathrm{P}(\YSZ|M)$ is the Gaussian probability distribution with which we have been working previously, and $\mathrm{P}(M)$ is the mass function. The number of clusters is a steep function of cluster mass, which (combined with the measuremtent uncertainty in \YSZ) results in relatively more low-mass than high-mass clusters at a given \YSZ, an effect commonly referred to as Eddington bias.

For our cluster sample we use the eighteen clusters from B13, fourteen of which have X-ray derived masses (see \S \ref{sec:Xray_obs}), and all of which have $\xi > 5$. For this analysis we use only the \onefifty\ data, the SPT band with the highest SZ sensitivity. For a list of cluster names, $\xi$ values, and redshifts for this sample, see Table \ref{tab:cluster_table}.

To fit for scaling relations we use a method similar to the one described in B13, which we modify to account for the cluster selection based on \YSZ\ instead of $\zeta$. In B13, we used a version of the CosmoMC \citep{lewis02b} analysis package, modified to include the cluster abundance likelihood in the CosmoMC likelihood calculation. All fitting is performed assuming a standard flat \LCDM \ cosmology, and using the WMAP 7-year data set. At each step in the chain, a point in the joint cosmological and scaling relation parameter space is selected. The Code for Anisotropies in the Microwave Background (CAMB) \citep{lewis00} is used to compute the matter power spectrum at twenty redshift bins between $0 < z < 2.5$, spaced logarithmically in $1+z$. The matter power spectrum, cosmological parameters, and \YSZ-\Mfh\ and \YX-\Mfh\ scaling relation parameters are then input to the cluster likelihood function. \YX\ is defined as the integrated X-ray flux within \rfh.

To calculate the cluster likelihood, first the matter power spectrum and cosmological parameters are used to calculate the cluster mass function, based on \citet{tinker08}. Next, the mass function is converted to the predicted cluster abundance in our observable space, $N(\YSZ,\YX,z)$. This conversion is accomplished using our standard \Ytheta-\Mfh\ scaling relation (equation \ref{eq:scaling_rel}), and the \YX-\Mfh\ scaling relation from B13:
\begin{equation}
\frac{\MX}{ 10^{14} \msun h^{-1} } = \left(\ax h^{3/2}\right) \left( \frac{\YX}{3 \times 10^{14} \msun \, {\rm keV}} \right)^{\bx} E(z)^{\cx},
\end{equation}
parametrized by the normalization factor \ax, the mass scaling \bx, the redshift evolution \cx, and the log-normal intrinsic scatter. This scaling relation is based on the relation used in \citet{vikhlinin09b}.

The predicted cluster density as a function of \YSZ, \YX, and $z$ can be written as follows:
\begin{equation}
  \frac{dN(\YSZ,\YX,z | \vec{p})}{d\YSZ \ d\YX \ dz} = \nonumber
\end{equation}
\begin{equation}
  \int P(\YSZ, \YX | M, z, \vec{p}) \ P(M, z | \vec{p}) \ \Phi(\YSZ) \ dM,
  \label{eq:grid}
\end{equation}
where $\vec{p}$ is the set of cosmological and scaling relation parameters, and $\Phi(\YSZ)$ is the selection function in \YSZ. This predicted cluster density function differs from B13 in that the selection function must be transformed from a Heaviside step function at $\xi=5$ into a function of \YSZ. We assume that \YSZ\ and $\xi$ can be related with a log-normal distributed scaling relation, and that the selection in B13 can therefore be well-approximated by an error-function in \YSZ. We then define our selection function as:
\begin{equation}
\Phi(\YSZ) = \frac{1}{2} \mathrm{erf} \left( \frac{ \YSZ - \Ythresh} {\sqrt{2} \ \Ythresh \ D} \right) + \frac{1}{2},
\end{equation}
where the selection threshold, \Ythresh \ is defined as the \YSZ\ value corresponding to $\xi = 5$ at the redshift $z$. We estimate \Ythresh\ by fitting a \YSZ-$\xi$ scaling relation of the form:
\begin{equation}
\YSZ = A \xi^B E(z)^C,
\end{equation}
using the catalog of SPT observed clusters given in R13. The width of the selection error-function is given by the scatter in the \YSZ-$\xi$ scaling relation, D.

We evaluate equation \ref{eq:grid} on a $200 \times 200 \times 30$ grid in (\YSZ, \YX, $z$) space, and convolve with a Gaussian error term in \YSZ\ to account for the measurement noise. The width of the Gaussian is given by the uncertainty in \YSZ\ as a function of \YSZ, $\delta \YSZ(\YSZ)$, as determined by the cluster parametrization MCMC (see \S \ref{sec:YSZ}).

The likelihood function of the observed cluster sample is defined by the Poisson probability:
\begin{eqnarray}
  \mathrm{Log} \left( \mathcal{L}(\vec{p}) \right) = \sum_i \mathrm{Log} \left( \frac{dN(\YSZ_i,\YX_i,z_i, | \vec{p})}{d\YSZ \ d\YX \ dz} \right) - \nonumber \\
 \int \frac{dN(\YSZ,\YX,z, | \vec{p})}{d\YSZ \ d\YX \ dz} \ d\YSZ \ d\YX \ dz,
  \label{eq:lnlike}
\end{eqnarray}
where the summation is over the SPT clusters in our catalog. Note also that this is the unnormalized log-likelihood. 

There is a complication, in that \YX\ is dependent on the cosmological parameters. $\YX \equiv M_\mathrm{g} \ T_\mathrm{X}$, where $M_\mathrm{g}$ is the gas mass within \rfh, and $T_\mathrm{X}$ is the core-excised X-ray temperature in an annulus between $0.15 \times \rfh$ and $1.0 \times \rfh$. To maintain consistency with the cosmological parameters, we recalculate \YX\ for each cluster at every step in CosmoMC, given the current \YX-\Mfh\ relation and \rfh.  In the likelihood, we add $\sum_i \mathrm{Log}(\YX_i)$ to the right hand side of equation \ref{eq:lnlike} to account for the recalculation of \YX. For a detailed explanation of this correction term, see Appendix B of B13.

To account for measurement error in \YX\ and $z$ for each cluster, we marginalize over the relevant parameter, weighted by a Gaussian likelihood determined by its uncertainty. For the few clusters without observed \YX\ data, we instead weight the marginalized parameter by a uniform distribution over the allowed parameter range.

The likelihood of this set of cosmological and scaling relation parameters is then used by CosmoMC in the acceptance/rejection computation.  
Only the \Ytheta-\Mfh\ scaling relation parameters are of interest to us in this analysis. The cosmological and \YX-\Mfh\ scaling relation parameters were used as a crosscheck to verify that the results were in agreement with the analysis performed on these clusters in B13, but will not be presented here. All parameters differed from the values presented in B13 by $<< 1\sigma$.

\subsection{\YSZ-\Mfh\ Scaling Relation Results}
\label{sec:B13_scaling_rel_results}

We use CosmoMC to fit \Ytheta-\Mfh\ scaling relations for a range of angular apertures, and find a broad minimum in scatter in the range $0.5'$ - $0.75'$, with a minimum intrinsic log-normal scatter of $21\pm11\%$. 
No priors are placed on these scaling relation parameters. 
The scatter in the $\zeta$-\Mfh\ scaling relation for these clusters is comparable, at $21\pm9\%$. 
The scaling relation parameters for the \Ysf-\Mfh\ scaling relation are given in Table \ref{tab:scaling_rels}. 

We also fit mass scaling relations for \Yrho\ integrated within a range of physical radii, $\rho$, from $0.1$ Mpc to $0.5$ Mpc.  
We find a broad minimum in scatter in the range $0.2$ - $0.3$ Mpc, with a minimum intrinsic log-normal scatter of $23\pm5\%$. 
This is comparable to the scatter in both the $\zeta$ and \Ysf\ mass scaling relations.
The parameters for the nominal \YthM mass scaling relation ($0.3$ Mpc corresponds to $0.75'$ at the survey median redshift of $z = 0.6$) are listed in Table \ref{tab:scaling_rels}.

\subsection{Cluster Masses}

To calculate the masses of the clusters, the \Ysf \ CosmoMC chains were used. The probability density function for the mass was computed on a grid for each step in the CosmoMC chains. These probability density functions were then combined to obtain a mass estimate fully marginalized over all scaling relation and cosmological parameters. This was done for CosmoMC chains containing only \Ysf \ data, and no \YX\ data, and vice versa, to obtain mass estimates based on only the SZ and X-ray data respectively. The cluster \Mfh \ masses derived from the \Ysf \ and \YX\ data (\MSZ \ and \MX \ respectively) can be found in Table \ref{tab:cluster_table}, along with the corresponding \Ysf \ and \YX\ values. \Ysf values are given in \msun keV for ease of comparison with \YX.

\begin{figure}
\includegraphics[width=\columnwidth]{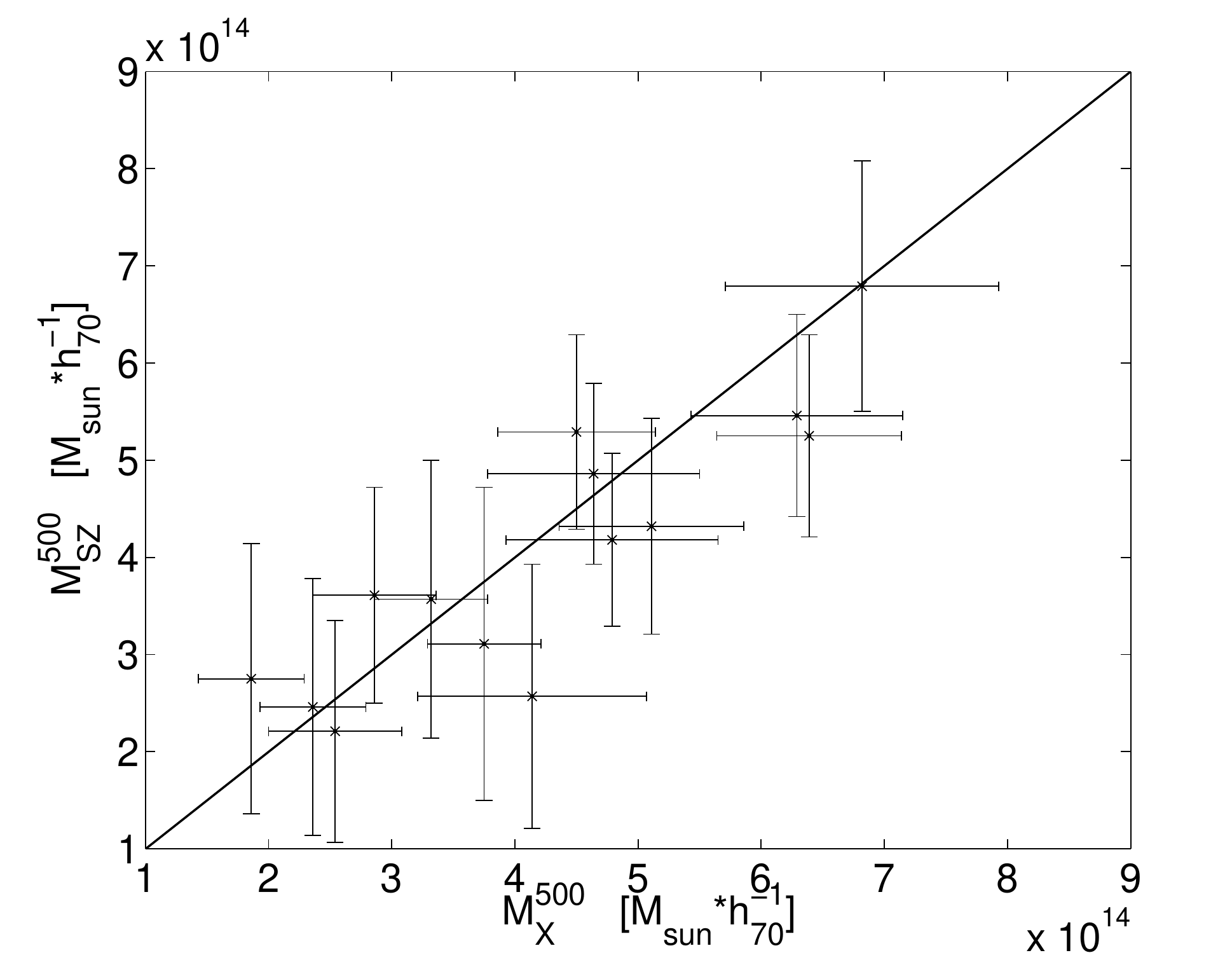}
\caption[]{Masses computed from \Ysf \ for the 14 SPT observed clusters in Table \ref{tab:cluster_table} versus corresponding \MX \ values. For reference we overplot the relation $\MSZ = \MX$.}
\label{fig:MSZvsMX}
\end{figure}

Figure \ref{fig:MSZvsMX} shows the cluster masses calculated from the \YSZ-\Mfh\ scaling relation versus the masses calculated from the \YX-\Mfh\ scaling relation for the B13 cluster sample. The solid line is the reference line $\MSZ = \MX$.

\begin{table*}
\begin{center}
\caption{\YSZ-$M$ Scaling Relation Parameters}
\begin{tabular}{l|ccccc|c}
  &  &  & MCMC & & & MF \\
\hline \hline
 Data Set & Integration & \asz & \bsz & \csz & Scatter    & Scatter \\
          & Radius      & ($\times 10^{-4}$) &      &      &    &  \\
\hline
\multirow{2}{*}{tSZ-Only S10 Sims} & $0.75'$ & $1.44\pm0.11$ & $1.20\pm0.11$ & $1.63\pm0.24$ & $23\pm2\%$ & \multirow{2}{*}{$27\pm2\%$} \\
& $0.3$ Mpc & $1.53\pm0.16$ & $1.26\pm0.17$ & $1.13\pm0.13$ & $28\pm2\%$ \\ \hline
\multirow{2}{*}{Full-Noise S10 Sims} & $0.75'$ & $1.37\pm0.10$ & $1.04\pm0.11$ & $1.02\pm0.20$ & $27\pm1\%$ & \multirow{2}{*}{$27\pm2\%$}\\
& $0.3$ Mpc & $1.49\pm0.18$ & $1.12\pm0.22$ & $0.53\pm0.25$ & $34\pm2\%$ \\ \hline
\multirow{2}{*}{B13 SPT Observed Clusters} & $0.75'$ & $1.85\pm0.36$ & $1.77\pm0.35$ & $0.96\pm0.50$ & $\ \ \ 21\pm11\%\tablenotemark{a}$  & \multirow{2}{*}{$\ \thinspace  21\pm9\%\tablenotemark{a}$} \\
& $0.3$ Mpc & $2.09\pm0.35$ & $1.43\pm0.20$ & $0.35\pm0.28$ & $\ \thinspace  26\pm9\%\tablenotemark{a}$ \\ \hline
\end{tabular}
\label{tab:scaling_rels}
\tablecomments{
The tSZ-only maps contain only thermal SZ signal, while the full-noise S10 maps include tSZ, CMB, point sources, atmospheric noise, and realistic SPT instrumental noise.
The values of scatter reported for the simulations are fractional scatter, while the values reported for the B13 clusters are intrinsic log-normal scatter.
For comparison with the scatter in each \YSZ-$M$ scaling relation we list the scatter in the corresponding MF derived $\zeta$-$M$ scaling relation for the same data set. 
}
\tablenotetext{a}{These values are intrinsic log-normal scatter.}
\end{center}
\end{table*}

\subsection{$\YSZ(\rfh$)}
\label{sec:y_five_hundred}

The self-similar model of cluster formation assumes that clusters scale in well-defined ways based on their mass, typically defined within physical radii proportional to the critical density of the universe at the cluster's redshift (e.g., \citet{kravtsov12}). For this reason, studies of the scaling relations of clusters typically measure physical observables defined by this physical radius, usually \rfh. In this section, we will calculate $\YSZ(\rfh)$, denoted \Yfh, for comparison with other published parameters for the clusters in B13.

We investigated a method for estimating \rfh\ from SZ data, as a way to measure \Yfh\ solely from SZ data. This method proved to be problematic however, because it required estimating \Mfh\ from a fixed angular aperture, and calculating \rfh\ from that estimate. This results in the scatter in the \Ysf-\Mfh\ scaling relation feeding back into the calculation of \Yfh. Instead, we use the X-ray determined \rfh\ in our calculations of \Yfh.

In Table \ref{tab:cluster_table}, we give the measured \Yfh\ values for our cluster sample.
We note that, as defined in equation \ref{eq:ysz_theta}, the MCMC fits for a cylindrically projected measure of \Yfh\, rather than the spherical de-projected value often used in other \YSZ-$M$ scaling relation results (e.g., A11, \citet{arnaud10}). \Yfh\ values are given in \msun keV here, for comparison with A11.

A11 describes a template fitting method of estimating \Yfh, which uses an SZ source template motivated from X-ray measurements of each cluster. The profile is assumed to match the product of the best-fit gas density profile to the X-ray measurements of each cluster, and the universal temperature profile of \citet{vikhlinin06}. These profiles are multiplied together to produce the radial pressure profile, and projected onto the sky using a line-of-sight integral through the cluster. A11 then constructs a spatial filter using equation \ref{eq:psi}, and this X-ray derived source model. The X-ray determined cluster position is used to place priors on the cluster location to prevent maximization bias in the recovered \Yfh \ values. \Yfh \ is calculated by integrating the source model over a solid angle corresponding to \rfh, as in equation \ref{eq:ysz_theta}.

\begin{figure}
\includegraphics[width=\columnwidth]{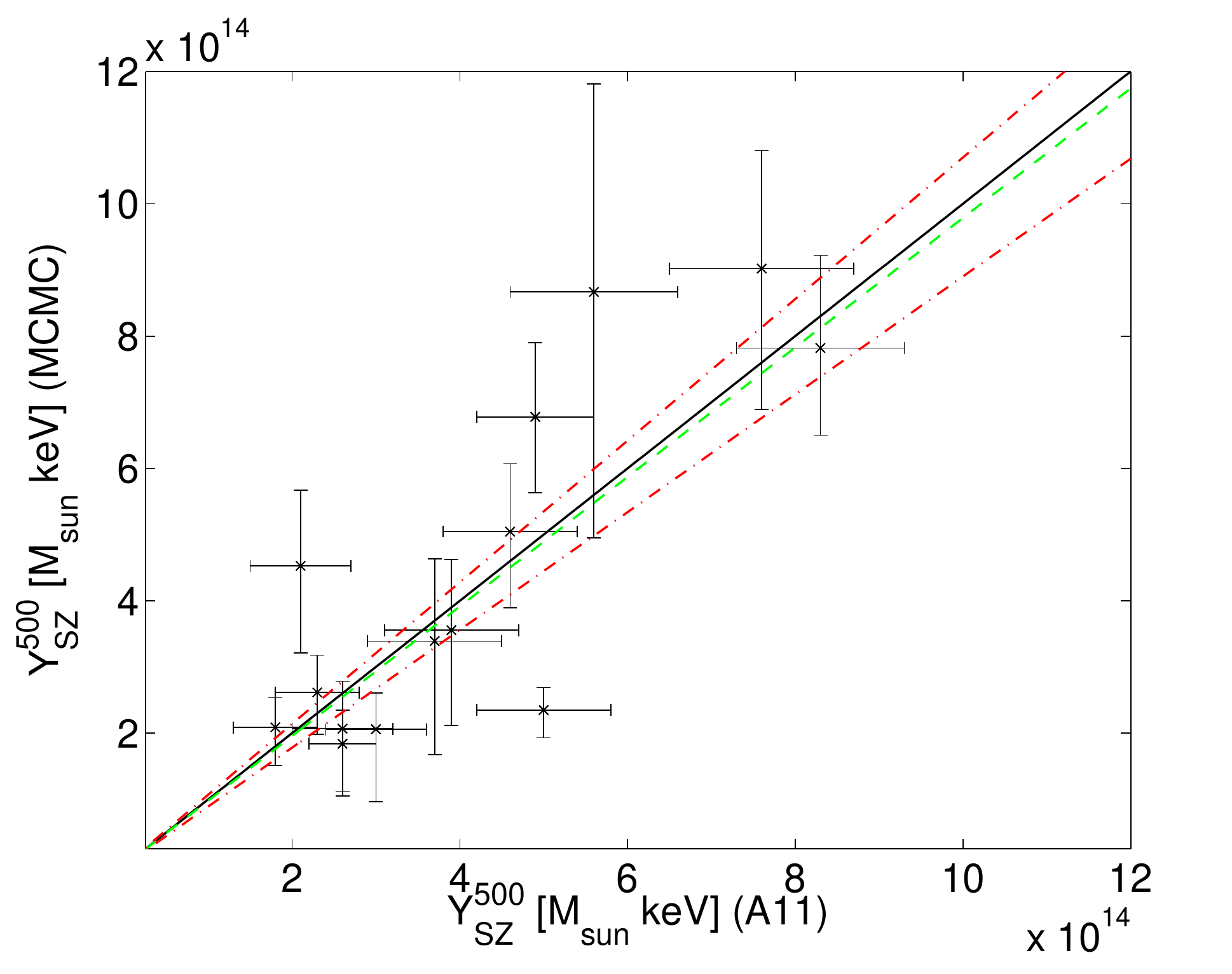}
\caption[]{\Yfh \ for the 14 SPT clusters from Table \ref{tab:cluster_table} calculated by the MCMC method described here, and by the MF method of \citet{andersson11}. We also show the reference line $\Yfh \mathrm{(MCMC)} = \Yfh \mathrm{(A11)}$ (solid), the best-fit line (green dashed), and the uncertainty in the fit defined as the range for which $\Delta \chi^2 < 1$ compared to the best-fit (red dot-dashed). The best-fit normalization is $A = 0.98 \pm 0.09$, demonstrating that the scaling relation is consistent with equality between $\Yfh \mathrm{(MCMC)}$ and $\Yfh \mathrm{(A11)}$.}
\label{fig:mcmc_vs_a11}
\end{figure}

In Figure \ref{fig:mcmc_vs_a11}, we plot the \Yfh \ estimated by the MCMC method against the \Yfh \ estimated by the template fitting method in A11. The best-fit relation between the two is $\Yfh(\mathrm{MCMC}) = (0.98\pm0.09) \ \Yfh(\mathrm{MF})$, where the uncertainty is the range for which $\Delta \chi^2 < 1$ compared to the best-fit. We see that these two methods of calculating \Yfh \ are consistent, that is, the best-fit scaling relation is consistent with equality between $\Yfh(\mathrm{MCMC})$ and $\Yfh(\mathrm{A11})$. 
The scatter about the expected one-to-one line here is dominated by differences in cluster model shape between the two methods (X-ray derived SZ profile versus $\beta$-model).

\begin{figure}
\includegraphics[width=\columnwidth]{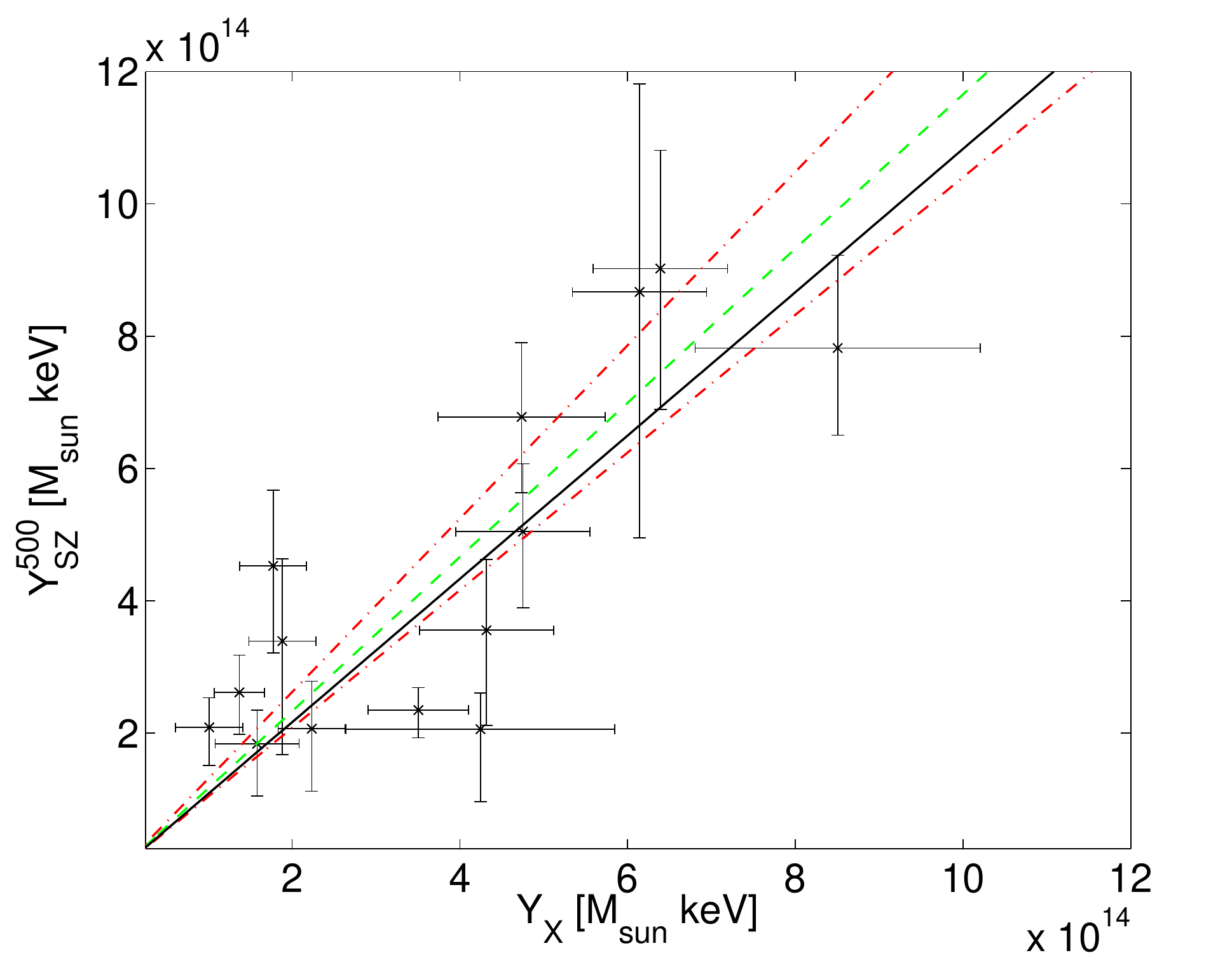}
\caption[]{\Yfh(MCMC) \ versus \YX\ for the 14 SPT clusters from Table \ref{tab:cluster_table}. We also show the expected scaling relation from \citet{arnaud10}: $\Yfh = 1.08 \ \YX$ (solid), the best-fit line (green dashed), and the uncertainty in the fit defined as the range for which $\Delta \chi^2 < 1$ compared to the best-fit (red dot-dashed). The best-fit normalization is $A = 1.17 \pm 0.12$, consistent with the expected scaling between \Yfh \ and \YX.}
\label{fig:ysz_vs_yx}
\end{figure}

We also verify that our \Yfh \ values for these clusters are in agreement with the \YX\ values presented in B13, given the expected \YSZ-\YX\ scaling. Figure \ref{fig:ysz_vs_yx} shows the \Yfh \ values of our catalog of SPT observed clusters plotted against their \YX\ values from B13. 

We can make a prediction of the relationship between \YSZ\ and \YX\ based on the universal pressure profile from \citet{arnaud10}, based on X-ray measurements of a representative sample of local, massive clusters.  Even though \YSZ\ and \YX\ are effectively measures of the cluster pressure, they depend on the details of the shape of the profile differently, which can still vary somewhat between clusters.  Assuming the \citet{arnaud10} pressure profile, we predict a relationship of $\Yfh = 1.08 \ \YX$, where \Yfh \ is integrated within a fixed angular aperture corresponding to \rfh \ (often called a cylindrical projection). In Figure \ref{fig:ysz_vs_yx}, we plot the \YSZ\ estimated by the MCMC method against the \YX\ measured in B13. We fit a scaling relation of the form $\Yfh = A \ \YX$, and find that the best-fit normalization is $A = 1.17\pm0.12$, consistent with the expected normalization. This fit has a total $\chi^2$ of $19.46$ for 14 degrees of freedom, with a probability to exceed of $P = 0.15$. The uncertainty in the normalization is the range for which $\Delta \chi^2 < 1$ compared to the best-fit.

\begin{table*}
\begin{center}
\caption{SPT Cluster Fluxes and Masses}
\begin{tabular}{lcccccc}
\hline \hline
Object Name & $z$ & \Ysf & \Yfh & \YX & \MSZ & $\MX$ \\
 & & ($10^{14} \msun \mathrm{keV}$) & ($10^{14} \msun \mathrm{keV}$) & ($10^{14} \msun \mathrm{keV}$) & ($10^{14} \msun h_{70}^{-1}$) & ($10^{14} \msun h_{70}^{-1}$) \\
\hline
SPT-CL J0509-5342 & $0.463$ & $0.9\pm0.1$ & $3.6^{+1.4}_{-1.1}$ & $4.3\pm0.8$ & $4.32\pm1.11$ & $5.11\pm0.75$  \\
SPT-CL J0511-5154\tablenotemark{a} & $0.74$ & $1.2\pm0.2$ & $ - $ & $ - $ & $2.79\pm1.43$ & $ - $ \\
SPT-CL J0521-5104\tablenotemark{a} & $0.72$ & $1.1\pm0.2$ & $ - $ & $ - $ & $2.46\pm1.32$ & $ - $ \\
SPT-CL J0528-5259 & $0.765$ & $1.1\pm0.2$ & $1.8^{+0.8}_{-0.5}$ & $1.6\pm0.5$ & $2.21\pm1.14$ & $2.54\pm0.54$  \\
SPT-CL J0533-5005 & $0.881$ & $1.4\pm0.2$ & $2.1^{+0.6}_{-0.4}$ & $1.0\pm0.4$ & $2.75\pm1.39$ & $1.86\pm0.43$  \\
SPT-CL J0539-5744\tablenotemark{a} & $0.77$ & $1.0\pm0.2$ & $ - $ & $ - $ & $1.93\pm0.93$ & $ - $ \\
SPT-CL J0546-5345 & $1.067$ & $2.0\pm0.3$ & $5.0^{+1.1}_{-1.0}$ & $4.8\pm0.8$ & $4.18\pm0.89$ & $4.79\pm0.86$  \\
SPT-CL J0551-5709 & $0.423$ & $0.7\pm0.1$ & $3.4^{+1.7}_{-1.2}$ & $1.9\pm0.4$ & $3.57\pm1.43$ & $3.32\pm0.46$  \\
SPT-CL J0559-5249 & $0.611$ & $1.6\pm0.2$ & $9.0^{+2.1}_{-1.8}$ & $6.4\pm0.8$ & $5.46\pm1.04$ & $6.29\pm0.86$  \\
SPT-CL J2301-5546\tablenotemark{a} & $0.748$ & $1.0\pm0.2$ & $ - $ & $- $ & $1.89\pm0.89$ & $ - $ \\
SPT-CL J2331-5051 & $0.571$ & $1.4\pm0.2$ & $2.3^{+0.4}_{-0.3}$ & $3.5\pm0.6$ & $5.29\pm1.00$ & $4.50\pm0.64$  \\
SPT-CL J2332-5358 & $0.403$ & $0.9^{+0.2}_{-0.1}$ & $8.7^{+3.7}_{-3.1}$ & $6.1\pm0.8$ & $5.25\pm1.04$ & $6.39\pm0.75$  \\
SPT-CL J2337-5942 & $0.781$ & $3.1\pm0.2$ & $7.8^{+1.3}_{-1.4}$ & $8.5\pm1.7$ & $6.67\pm1.29$ & $6.82\pm1.11$  \\
SPT-CL J2341-5119 & $0.998$ & $2.3\pm0.2$ & $6.8\pm1.1$ & $4.7\pm1.0$ & $4.86\pm0.93$ & $4.64\pm0.86$  \\
SPT-CL J2342-5411 & $1.074$ & $1.5\pm0.3$ & $2.6\pm0.6$ & $1.4\pm0.3$ & $2.46\pm1.32$ & $2.36\pm0.43$  \\
SPT-CL J2355-5056 & $0.320$ & $0.4\pm0.1$ & $2.1^{+0.9}_{-0.7}$ & $2.2\pm0.4$ & $3.11\pm1.61$ & $3.75\pm0.46$  \\
SPT-CL J2359-5009 & $0.774$ & $1.4\pm0.2$ & $4.5^{+1.3}_{-1.1}$ & $1.8\pm0.4$ & $3.61\pm1.11$ & $2.86\pm0.50$  \\
SPT-CL J0000-5748 & $0.701$ & $1.1\pm0.2$ & $2.1^{+1.1}_{-0.6}$ & $4.2\pm1.6$ & $2.57\pm1.36$ & $4.14\pm0.93$  \\
\hline
\end{tabular}
\label{tab:cluster_table}
\tablecomments{
Cluster redshifts and X-ray fluxes are quoted from \citet{benson13}. 
\Ysf \ is the integrated Comptonization within $0.75'$, calculated with our \YSZ\ MCMC method. 
\Yfh \ is the integrated Comptonization within \rfh. 
\Ysf\ and \Yfh\ values are given in \msun keV for comparison to \YX\ and the \YSZ\ values from A11. 
\Ysf\ and \Yfh\ are cylindrically projected. 
\MSZ \ and \MX \ are estimates of \Mfh \ calculated from the same CosmoMC chains, using only the \Ysf \ and \YX\ data respectively.
}
\tablenotetext{a}{These clusters have only SZ data, and no X-ray observations.}
\end{center}
\end{table*}

\section{Conclusions}
\label{sec:conc}

We describe and implement a method of constraining \YSZ\ generalizable to any cluster profile, and we show that this method accurately recovers \YSZ\ in simulations. 
We compare \YSZ\ to SPT cluster detection significance, focusing on scatter with mass.  
Finally, we apply this method to clusters detected in the SPT-SZ survey, and compare the estimated \YSZ\ values to \YSZ\ estimated by a template fitting method, and to \YX.

We apply our method to clusters in simulated tSZ-only maps and measure \Ytheta, the integrated Comptonization within a constant angular aperture.  
We find that \YSZ\ is measured with the lowest fractional scatter in an aperture comparable to the SPT beam size ($\aprx1'$ FWHM at \onefifty).
We fit \Ytheta-\Mvir\ scaling relations for a range of angular apertures and find a minimum fractional scatter of $23\pm2\%$ in \YSZ, at a fixed mass, with the minimum occurring for an angular aperture of $0.75'$.
We also calculate \YSZ\ within a range of physical radii, $\rho$, and find a minimum scatter in \Yrho\ at an integration radius of $0.3$ Mpc, which corresponds roughly to $0.75'$ at the survey median redshift ($z = 0.6$), with a fractional scatter of $28\pm2\%$ at a fixed mass.
Using the same simulated clusters, we also fit a $\zeta$-\Mvir\ relation, where $\zeta$ is the matched filter SZ detection significance measured by SPT, and find a fractional scatter of $27\pm2\%$.

We also analyze clusters in simulations including tSZ, CMB, point sources, atmospheric noise, and realistic SPT instrumental noise. 
In these full-noise simulations, the \Ysf-\Mvir\ scaling relation has $27\pm1\%$ scatter, the \YthM-\Mvir\ scaling relation has $34\pm2\%$ scatter, and $\zeta$-\Mvir\ scaling relation has $27\pm2\%$ scatter. 
These simulations demonstrate that scatter in \Ytheta\ is comparable to the scatter in $\zeta$.

To investigate \YSZ\ scaling relations in SPT observed clusters, we fit \Ytheta-\Mfh\ and \Yrho-\Mfh\ scaling relations to the sample of eighteen SPT clusters described and examined in \citet{benson13}. 
Of these, fourteen have X-ray observations and measured \YX\ values, which we use to estimate the cluster \Mfh\ masses.
We fit the scaling relations using a version of CosmoMC, similar to the one described in \citet{benson13}, modified to account for the cluster selection based on \YSZ\ instead of SPT significance. 
For these clusters, the \Ysf-\Mfh\ scaling relation is found to have $21\pm11\%$ intrinsic log-normal scatter in \YSZ\ at a fixed mass, the \YthM-\Mfh\ scaling relation has $26\pm9\%$ scatter, and the $\zeta$-\Mfh\ relation has $21\pm9\%$ scatter. 

We also calculate a cylindrically projected \Yfh, the integrated Comptonization within \rfh, for the clusters in the \citet{benson13} sample. 
We compare the \Yfh \ values recovered by our Markov-Chain Monte Carlo method to those calculated for the same clusters by the template fitting method described in A11 and find the two methods to be consistent. 
We further compare the MCMC derived \Yfh \ values to the \YX\ values for these clusters from \citet{benson13} and find that they are consistent with the expected scaling between \YSZ\ and \YX, based on the universal pressure profile of \citet{arnaud10}.

We have demonstrated, with both simulations with realistic SPT noise and SPT observed clusters, that \YSZ\ is most accurately determined in an aperture comparable to the SPT beam size. 
We have used this information in measuring \YSZ\ for the catalog of clusters observed with the SPT in the 2008 and 2009 seasons \citep{reichardt13}.
The SPT-SZ survey of $2500 \ \sqdeg$ was completed in November 2011, and has detected $\aprx 500$ clusters with a median redshift of $\aprx 0.5$ and a median mass of $\Mfh \ \aprx \ 2.3 \times 10^{14} \msun h^{-1}$.
The methods and results presented here will inform the measurement and use of \YSZ\ for the clusters detected in the full SPT-SZ survey.

\section*{Acknowledgments}

The South Pole Telescope program is supported by the National Science
Foundation through grant ANT-0638937.  Partial support is also
provided by the NSF Physics Frontier Center grant PHY-0114422 to the
Kavli Institute of Cosmological Physics at the University of Chicago,
the Kavli Foundation, and the Gordon and Betty Moore Foundation.  

We acknowledge the use of the Legacy Archive for
Microwave Background Data Analysis (LAMBDA).  Support for LAMBDA is
provided by the NASA Office of Space Science.  Galaxy cluster research
at Harvard is supported by NSF grant AST-1009012.  Galaxy cluster
research at SAO is supported in part by NSF grants AST-1009649 and
MRI-0723073.  The McGill group acknowledges funding from the National
Sciences and Engineering Research Council of Canada, Canada Research
Chairs program, and the Canadian Institute for Advanced Research.
The Munich group was supported by The Cluster of
Excellence ``Origin and Structure of the Universe'', funded by the
Excellence Initiative of the Federal Government of Germany, EXC
project number 153. R.J.F.\ is supported by a Clay Fellowship, 
and B.A.B.\ is supported by a KICP Fellowship. 
A. P.\ is supported by an NSF Graduate Research Fellowship under Grant No. DGE-1144152.
J.H.L.\ is supported by NASA through the Einstein Fellowship Program under Grant No. PF2-130094.
M.M.\ acknowledges support provided by
NASA through a Hubble Fellowship grant from STScI.
M.D.\ acknowledges support from an Alfred P.\ Sloan Research Fellowship, 
W.F.\ and C.J.\ acknowledge support from the Smithsonian Institution, 
and B.S.\ acknowledges support from the Brinson Foundation.

{\it Facilities:}
\facility{Blanco (MOSAIC)},
\facility{CXO (ACIS)},
\facility{Gemini-S (GMOS)},
\facility{Magellan:Baade (IMACS)},
\facility{Magellan:Clay (LDSS3)},
\facility{South Pole Telescope},
\facility{XMM-Newton (EPIC)}

\bibliography{saliwanchik2013}

\end{document}